%% file: main.tex
\newif\ifarxiv\arxivtrue%

\ifarxiv%
\pdfoutput=1
\fi

\documentclass[runningheads,orivec,envcountsect,envcountsame]{llncs}
\usepackage{silence}

\RequirePackage[paperheight=235mm,paperwidth=155mm,textwidth=12.2cm,textheight=19.3cm,inner=55pt]{geometry}

\ifarxiv%
\makeatletter
\def\@citecolor{blue}%
\def\@urlcolor{blue}%
\def\@linkcolor{blue}%

\def\orcidID#1{\href{http://orcid.org/#1}{\protect\raisebox{-1.25pt}{\protect\includegraphics{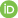}}}}
\makeatother
\fi%

\usepackage{microtype}
\usepackage[T1]{fontenc}
\usepackage{graphicx}

\usepackage{etoolbox}
\apptocmd{\sloppy}{\hbadness 10000\relax}{}{} 

\usepackage{mathpartir}

\usepackage[hidelinks]{hyperref}

\usepackage{tikz}
\usetikzlibrary{
    cd,
    fit,
    calc,
    positioning,
    arrows,
    automata,
    shapes
}
\tikzset{
    baseline = (current bounding box.center),
    every state/.append style = {
		inner sep = 2pt,
		minimum size = 1.5em,
		initial text = {}
	},
	every edge/.append style = {
		->,
		>=stealth,
		bend angle=10,
		thick
	}
}

\usepackage{amsmath}
\usepackage{amssymb}
\usepackage{thmtools}

\newcommand{\id}{\mathrm{id}}
\newcommand{\Id}{\mathrm{Id}}
\newcommand{\ar}{\mathrm{ar}}

\newcommand{\N}{\mathbb{N}}

\newcommand{\Var}{\mathit{Var}}
\newcommand{\StExp}{\mathit{StExp}}
\newcommand{\Exp}{\mathit{Exp}}

\newcommand{\At}{\mathit{At}}
\newcommand{\Act}{\mathit{Act}}

\newcommand{\T}{\mathsf{T}}
\newcommand{\SL}{\mathsf{SL}}
\newcommand{\GA}{\mathsf{GA}}
\newcommand{\GC}{\mathsf{GC}}
\newcommand{\CA}{\mathsf{CA}}
\newcommand{\CS}{\mathsf{CS}}

\newcommand{\Set}{\mathbf{Set}}
\newcommand{\Coalg}{\operatorname{Coalg}}
\newcommand{\conv}{\operatorname{conv}}

\renewcommand{\P}{\mathcal{P}}
\newcommand{\D}{\mathcal{D}}
\newcommand{\R}{\mathcal{R}}

\newcommand{\supp}{\mathsf{supp}}

\newcommand{\into}{\hookrightarrow}

\newcommand{\bisim}{\mathrel{\raisebox{1pt}{\(\underline{\leftrightarrow}\)}}}

\newcommand{\diredge}{\curvearrowright}
\newcommand{\tr}[1]{\mathrel{\raisebox{-2pt}{\(\xrightarrow{#1}\)}}}
\newcommand{\etr}[1]{\mathrel{
    \raisebox{-2pt}{
        \color{blue}
        \(\xrightarrow{{\color{black}#1}}_{e}\)
    }
}}
\newcommand{\btr}[1]{\mathrel{
    \raisebox{-2pt}{
        \(\xrightarrow{{#1}}_{\mathsf b}\)
    }
}}
\newcommand{\eo}{\mathrel{\color{blue}\to_{\mathsf{e}}}}
\newcommand{\bo}{\mathrel{\to_{\mathsf{b}}}}

\usepackage{hyperref}
\usepackage[capitalise]{cleveref}

\spnewtheorem{instantiation}{Instantiation}{\bf}{\rmfamily} 
\crefformat{instantiation}{#2instantiation #1#3}
\Crefformat{instantiation}{#2Instantiation #1#3}

\begin{document}

\title{A General Completeness Theorem \texorpdfstring{\\}{}for Skip-free Star Algebras}

\author{%
    Tobias Kapp\'{e}\inst{1}\orcidID{0000-0002-6068-880X} \and 
    Todd Schmid\inst{2}\orcidID{0000-0002-9838-2363} 
}

\authorrunning{T.~Kappé and T.~Schmid}

\institute{%
    LIACS, Leiden University \\
    \email{t.w.j.kappe@liacs.leidenuniv.nl}
    \and
    Bucknell University \\
    \email{t.schmid@bucknell.edu}
}

\maketitle

\begin{abstract}
We consider process algebras with branching parametrized by an equational theory $\T$, and show that it is possible to axiomatize bisimilarity under certain conditions on $\T$.
Our proof abstracts an earlier argument due to Grabmayer and Fokkink (LICS'20), and yields new completeness theorems for skip-free process algebras with probabilistic (guarded) branching, while also covering existing completeness results.
\end{abstract}

\section{Introduction}
Regular expressions generated by a set of action symbols \(\Act\) are classically interpreted as regular languages, i.e., sets of words over \(\Act\) obtained by the usual union, concatenation, and Kleene star operations~\cite{kleene-56}.
Inspired by the introduction of process algebra as a formalization of communicating and concurrent processes, Milner gave an interpretation of regular expressions in terms of nondeterministic machine \emph{behaviours}, i.e., up to bisimilarity~\cite{milner-84}.
The labelled transition system obtained from a regular expression is constructed by interpreting the union operation as nondeterministic choice instead of language union and the Kleene star operation as a (finite) loop instead of repeated language concatenation.
Although introduced many years apart, Milner's interpretation of regular expressions is now known to be equivalent to Antimirov's derivative construction~\cite{antimirov-95}.

Milner transformed Salomaa's axioms for language equivalence of regular expressions~\cite{salomaa-66} into sound axioms for his behavioural semantics, and left completeness as an open problem.
This problem was only recently solved by Grabmayer~\cite{grabmayer-2022}, who was able to reduce the problem to results in a prior collaborative work with Fokkink~\cite{grabmayer-fokkink-2020}.
In op.\ cit., the authors prove that Milner's axioms are complete when restricted to a subset of regular expressions that they call \emph{1-free star expressions}, which are regular expressions in which $1$ (the unit for concatenation) does not appear, and the Kleene star is replaced by a binary star operation \(r_1 * r_2\).

Around the same time, Smolka et al.~\cite{smolkaFHKKS-20}\ introduced \emph{guarded Kleene algebra with tests} (or \emph{GKAT}) for reasoning about simple imperative programs.
Essentially, GKAT is a restriction of \emph{Kleene algebra with tests} (or \emph{KAT})~\cite{kozen-97} to programs constructed out of \textsf{if-then-else} and \textsf{while}.%
\footnote{
    The syntax and significance of this fragment of KAT was already noticed by Kozen and Tseng in~\cite{kozen-tseng-08}.
    The innovations of~\cite{smolkaFHKKS-20} are the axiomatization and complexity results, as well as a number of different interpretations of GKAT\@.
}
In~\cite{smolkaFHKKS-20}, the authors proposed an infinitary axiomatization of equivalence between GKAT programs.
They left open whether a finitary version of these axioms, similar to Milner's, is also complete.

More recently, a step towards proving completeness of GKAT was taken in~\cite{kappe-schmid-silva-2023}.
In that paper, it was shown that the finitary axiomatization of GKAT proposed in~\cite{smolkaFHKKS-20} is complete when restricted to GKAT programs that are 1-free in a sense similar to 1-free star expressions.
The program that acts like \(1\) in GKAT is \textsf{skip}, so the authors of~\cite{kappe-schmid-silva-2023} refer to these programs as \emph{skip-free} GKAT programs.
To achieve this result, it was necessary to analyze some tricky steps in the original completeness proof from~\cite{grabmayer-fokkink-2020}, and adapt them to the setting of skip-free GKAT\@.

In this paper, we study a unifying generalization of 1-free star expressions and skip-free GKAT programs that we call \emph{skip-free unified-star expressions}.
This generalization is based on the observation in~\cite{schmid-rozowski-rot-silva-22,schmid-thesis} that 1-free star expressions and skip-free GKAT programs can be seen as instances of process algebras parametrized by \emph{equational theories}, which capture computational effects such as nondeterminism and probability via their corresponding monads~\cite{plotkin-power-01,plotkin-power-02}.

Our main contribution is a complete axiomatization for each process algebra in this class, provided the equational theory satisfies two conditions that we refer to as \emph{admitting a support} and \emph{malleability}.
These properties essentially allow us to employ the strategy used by Grabmayer and Fokkink in~\cite{grabmayer-fokkink-2020}, but in a much more abstract setting.
The equational theories that correspond to 1-free star expressions and skip-free GKAT admit a support and are malleable, and so our abstract completeness theorem generalizes the results from~\cite{grabmayer-fokkink-2020,kappe-schmid-silva-2023}.
We furthermore obtain several new completeness theorems for bisimilarity semantics of process algebras considered in the literature, including one for the skip-free fragment of probabilistic regular expressions studied in~\cite{rozowskiS24} and another for a skip-free variation of the probabilistic version of GKAT studied in~\cite{rozowski-etal-2023}.

We assume the reader is familiar with the basics of category theory, i.e., functors, universal properties, and natural transformations (see, for example,~\cite{riehl-2016}).
\ifarxiv
    Omitted proofs can be found in the appendix.
\else
    Omitted proofs can be found in the full version of the paper~\cite{fullversion}.
\fi

\section{1-free Star Expressions}
To set the stage, we begin with a brief description of Grabmayer and Fokkink's completeness theorem and its proof technique~\cite{grabmayer-fokkink-2020}.
Fix a set of \emph{action symbols} \(\Act\).
The set of \emph{1-free star expressions} \(\StExp\) is generated by the grammar below.
\begin{equation}
    \label{def:1-free regex}
    r,r_1,r_2 ::= 0 \mid a \in \Act \mid r_1 + r_2 \mid r_1\cdot r_2 \mid r_1*r_2
\end{equation}
The operators $+$ and $\cdot$ are to be interpreted as nondeterministic choice and sequential composition, respectively.
The expression \(r_1*r_2\) is intended to mean ``run $r_1$ some number of times, and then run $r_2$'' (cf.\ the regular expression $r_1^* r_2$).
In the absence of parentheses, $*$ takes precedence over $\cdot$, which is evaluated before $+$.
So, \(a + b\cdot c * d\) should be read as \(a + (b \cdot (c * d))\).
We typically write \(r_1r_2\) instead of \(r_1 \cdot r_2\).
The process semantics for 1-free star expressions can be phrased in terms of a variant of labelled transition systems called \emph{charts}.

\begin{definition}%
    \label{def:chart}
    A \emph{chart} is a pair \((X, \delta)\) consisting of a set of \emph{states} \(X\) and a \emph{transition function} \(\delta \colon X \to \P(\Act \times (\checkmark + X))\).
    Here, \(\P\) is the finite powerset functor, \(\checkmark\) denotes the set \(\{\checkmark\}\), and \(+\) is disjoint union.
    Given \(x,y \in X\) and \(a \in \Act\), we write \(x \tr{a}_\delta y\) if \((a, y) \in \delta(x)\), and \(x \tr{a}_\delta \checkmark\) if \((a, \checkmark) \in \delta(x)\).
\end{definition}

Immediately we see that \(\delta\) can be either specified as a function or as a set of \emph{transition relations} \({\tr{a}}_\delta \subseteq X \times (\checkmark + X)\), one for each \(a \in \Act\).
We usually drop subscripts if the transition function can be inferred from context.
We write \({\to}\) for the union of the transition relations and say that \(x\) \emph{transitions} to \(y\) if \(x \to y\).

Given \((X, \delta)\) and \(U \subseteq X\), define \(\langle U \rangle_\delta = \{y \mid \exists x \in U, x \to_\delta^* y\}\).
The \emph{subchart of $(X, \delta)$ generated by U} is $(\langle U\rangle_\delta, \delta_U)$, where $\delta_U: \langle U \rangle_\delta \to \P(\Act \times (\checkmark + \langle U \rangle_\delta))$ is simply $\delta$ restricted to $\langle U \rangle_\delta$; we also write $\langle U \rangle_\delta$ for this chart.
When $x \in X$, we may write $\langle x \rangle_\delta$ for $\langle\{x\}\rangle_\delta$, and speak of \emph{the subchart of $(X, \delta)$ generated by $x$}.

The set of 1-free star expressions itself carries a chart structure, whose transitions are derived inductively from the inference rules in \cref{fig:syntactic chart}, in a way that is reminiscent of Antimirov's automaton construction~\cite{antimirov-95}.
We use \((\StExp, \delta)\) to denote this chart structure, and refer to it as the \emph{syntactic chart}.

\begin{figure}[t]
    \begin{gather*}
        \infer{~}{a \tr{a} \checkmark}
        \qquad
        \infer{r_1 \tr{a} \xi}{r_1 + r_2 \tr{a} \xi}
        \qquad
        \infer{r_2 \tr{a} \xi}{r_1 + r_2 \tr{a} \xi}
        \qquad
        \infer{r_1 \tr{a} s}{r_1r_2 \tr{a} sr_2}
        \qquad
        \infer{r_1 \tr{a} \checkmark}{r_1r_2 \tr{a} r_2}
        \\
        \infer{r_1 \tr{a} s}{r_1*r_2 \tr{a} s(r_1 * r_2)}
        \qquad
        \infer{r_1 \tr{a} \checkmark}{r_1*r_2 \tr{a} r_1 * r_2}
        \qquad
        \infer{r_2 \tr{a} s}{r_1*r_2 \tr{a} s}
    \end{gather*}
    \vspace*{-2em}
    \caption{
        Rules defining the transition structure of the syntactic chart \((\StExp, \delta)\).
        In the above, \(a \in \Act\), \(r_1,r_2,s \in \StExp\), and \(\xi \in \checkmark + \StExp\).
    }%
    \label{fig:syntactic chart}
\end{figure}

Every 1-free star expression \(r\) generates a finite subchart \(\langle r\rangle\) of \((\StExp, \delta)\): the \emph{syntactic chart} of \(r\).
The equivalence for 1-free star expressions that Grabmayer, Fokkink, and Milner sought to axiomatize is \emph{bisimilarity}, defined as follows.

\begin{definition}%
    \label{def:chart bisimilarity}
    A \emph{bisimulation} between charts \((X, \delta_X)\) and \((Y, \delta_Y)\) is a relation \(R \subseteq X \times Y\) such that for any \((x,y) \in R\) and \(a \in \Act\),
    (1) \(x \tr{a} \checkmark\) if and only if \(y \tr{a} \checkmark\),
    (2) if \(x \tr{a} x'\), then there is a \((x',y') \in R\) such that \(y \tr{a} y'\), and
    (3) if \(y \tr{a} y'\), then there is a \((x',y') \in R\) such that \(x \tr{a} x'\).
    We write \(x \bisim y\) and say that \(x\) and \(y\) are \emph{bisimilar} if \((x, y) \in R\) for some bisimulation \(R\).
\end{definition}

Note that the relation \(\bisim\) is the largest bisimulation between two charts.
Furthermore, $\bisim$ is an equivalence relation when restricted to a single chart.
Another notion that we will rely on is that of a \emph{chart homomorphism}.

\begin{definition}%
    \label{def:chart homomorphism}
    A \emph{homomorphism} of charts \(h \colon (X, \delta_X) \to (Y, \delta_Y)\) is a function \(h \colon X \to Y\) such that the graph of \(h\) is a bisimulation.
\end{definition}

It is straightforward to show that charts and chart homomorphisms form a category, i.e., identity maps \(\id \colon (X, \delta) \to (X, \delta)\) are homomorphisms and any composition of homomorphisms is a homomorphism.
Furthermore, bisimilarity can be characterized using chart homomorphisms, via the following lemma.

\begin{lemma}%
    \label{lem:characterize chart bisim}
    Given states \(x \in X\) and \(y \in Y\) of charts \((X, \delta_X)\) and \((Y, \delta_Y)\), \(x \bisim y\) if and only if there is a third chart \((Z, \delta_Z)\) and chart homomorphisms \(h \colon (X, \delta_X) \to (Z, \delta_Z)\) and \(k \colon (X, \delta_X) \to (Z, \delta_Z)\) such that \(h(x) = k(y)\).
\end{lemma}

This characterization is useful in several ways; for one, it connects the existing notion of bisimilarity of charts to the more abstract definition in the next section.
Another consequence of \cref{lem:characterize chart bisim} is that for charts \((U, \delta_U)\) and \((X, \delta)\) with \(U \subseteq X\), \((U, \delta_U)\) is a subchart of \((X, \delta)\) generated by $U$ if and only if the inclusion map \(U \into X\) is a chart homomorphism.
Furthermore, for any \(r,s \in \StExp\), \(r \bisim s\) as states of \((\StExp, \delta)\) iff they are bisimilar as states of \(\langle r \rangle\) and \(\langle s\rangle\) respectively.

\paragraph*{Axiomatization.}
Milner showed that bisimilarity is a congruence with respect to the regular expression operations~\cite{milner-84}.
This led him to give axioms for bisimilarity using a variation on Salomaa's axioms for language equivalence of regular expressions~\cite{salomaa-66}.
Grabmayer and Fokkink adapted these axioms for 1-free regular expressions in~\cite{grabmayer-fokkink-2020}, which we recall in \cref{fig:regex axioms}.%
\footnote{
    Grabmayer and Fokkink included the additional equation \((x*y)z = x*(yz)\) in their axiomatization, but this is derivable from the other axioms.
}
For reasons that will become clear in \cref{sec:M-systems}, we write \(\SL^*\) to denote the theory in the figure.

\begin{figure}[t]
    \begin{gather*}
        \begin{aligned}
            x + y &= y + x \\
            (x + y) + z &= x + (y + z) \\
            x + x &= x = x + 0
        \end{aligned}
        \quad
        \begin{aligned}
            (x + y)z &= xz + yz \\
            (xy)z &= x(yz) \\
            0x &= 0
        \end{aligned}
        \quad
        \begin{gathered}
            \begin{aligned}
                x*y &= x(x*y) + y
            \end{aligned}\\
            \infer{x = yx + z}{x = y*z}
        \end{gathered}
    \end{gather*}
    \vspace*{-2em}
    \caption{
        The axioms proposed by Grabmayer and Fokkink in~\cite{grabmayer-fokkink-2020}.
        The theory \(\SL^*\) consists of the axioms above and equational logic (not pictured above).
    }%
    \label{fig:regex axioms}
\end{figure}

\begin{definition}%
    \label{def:regex prov equiv}
    Given \(r_1,r_2 \in \StExp\), we write \(\SL^* \vdash r_1 = r_2\) if there is a derivation of the equation \(r_1 = r_2\) from the axioms of \(\SL^*\).
\end{definition}

The following theorem was the main result in~\cite{grabmayer-fokkink-2020}.

\begin{theorem}[Soundness and completeness of $\SL^*$]%
    \label{thm:gf completeness}
    Let \(r_1, r_2 \in \StExp\).
    Then \(\SL^* \vdash r_1 = r_2\) if and only if \(r_1 \bisim r_2\) as states in \((\StExp, \delta)\).
\end{theorem}

The forward direction is called \emph{soundness}, and it can be proven by induction on derivations.
The reverse direction is \emph{completeness}, and requires a more involved proof with several steps.
To arrive at the completeness result, some mathematical machinery has to be developed, and two particularly treacherous proofs have to be worked out.
In the remainder of this section, we give an overview of the necessary techniques as preparation for our abstract approach in the sequel.

\paragraph*{The Completeness Proof.}
The first step on our journey is to cast a chart as a system of equations and to study its solutions.
Formally, given a chart \((X, \delta)\), we treat each state \(x \in X\) as an unknown, and add the formal equation \(x = a_1 x_1 + \cdots + a_n x_n + b_1 + \cdots + b_m\), with the right-hand side determined by the transition function at \(x\), \(\delta(x) = \{(a_1, x_1), \dots, (a_n, x_n), (b_1, \checkmark), \dots, (b_m, \checkmark)\}\).

A solution to the system of equations for \((X, \delta)\) maps states to expressions in a way that satisfies the equations, up to equivalence.
To formalize this, we need some notation:
for a finite set of 1-free star expressions \(S = \{r_1, \dots, r_n\}\), we write \(\sum_{r \in S} r\) to denote \(r_1 + \cdots + r_n\).
This is well-defined up to equivalence, because the axioms in \cref{fig:regex axioms} include commutativity and associativity.
If \(S = \{r \in \StExp \mid P(r)\}\) for some finite predicate \(P\), we write \(\sum_{P(r)} r\) instead of \(\sum_{r \in S} r\).

\begin{definition}%
    \label{def:regex solution}
    A \emph{solution} to a chart \((X, \delta)\) is a map \(\phi \colon X \to \StExp\) such that for any \(x \in X\), \(\SL^* \vdash \phi(x) = \sum_{x \tr{a} y} a\phi(y) + \sum_{x \tr{b} \checkmark} b\).
    Two solutions \(\phi,\psi\) are \emph{equivalent} if for any \(x \in X\), \(\SL^* \vdash\phi(x) = \psi(x)\).
    A chart \((X, \delta)\) is said to have a \emph{unique} solution if it has exactly one solution up to equivalence.
\end{definition}


\begin{example}%
    \label{example:solution}
    Let $X = \{ x, x' \}$ and suppose $(X, \delta_X)$ is the chart
    below.
    \[
    \begin{tikzpicture}
        \node[state] (x) {$x$};
        \node[state, right=2cm of x] (x') {$x'$};
        \node (check) at ($(x)!0.5!(x')$) {$\checkmark$};
        \draw[->] (x) to[bend right] node[below] {$a$} (x');
        \draw[->] (x) to node[above] {$b$} (check);
        \draw[->] (x') to[loop right] node[right] {$c$} (x');
        \draw[->] (x') to node[above] {$d$} (check);
    \end{tikzpicture}
    \]
    A solution to $(X, \delta_X)$ is a function $\phi: X \to \StExp$ such that
    \(
    \SL^* \vdash \phi(x) = a \cdot \phi(x') + b
    \)
    and
    \(
    \SL^* \vdash \phi(x') = c \cdot \phi(x') + d
    \).
    The trick to solving this system is to apply the last axiom in \Cref{fig:regex axioms} to the second equation, and choose $\phi(x') = c*d$.
    If we also set $\phi(x) = b + a(c*d)$, then $\phi$ is a solution by construction.
\end{example}

The following property corresponds to~\cite[Lemma~2.2]{schmid-rot-silva-21} and~\cite[Theorem~2.2]{schmid-rot-silva-21}, which characterize solutions to charts as chart homomorphisms into the quotient of \(\StExp\) by provable equivalence.
Given \(r_1,r_2 \in \StExp\), write \(r_1 \equiv r_2\) to denote that \(\SL^* \vdash r_1 = r_2\), and write \([r_1]_\equiv\) to denote the \(\equiv\)-equivalence class of \(r_1\).

\begin{proposition}%
    \label{prop:regex mod bisim}
    There is a unique \([\delta]_\equiv \colon \StExp/{\equiv} \to \P(\Act \times (\checkmark + \StExp))\) such that the quotient \([-]_\equiv \colon \StExp \to \StExp/{\equiv}\) is a chart homomorphism from \((\StExp, \delta)\) to \((\StExp/{\equiv}, [\delta]_\equiv)\).
    Moreover, \(\phi \colon X \to \StExp\) is a solution to a chart \((X, \delta)\) if and only if \([-]_\equiv \circ \phi \colon (X, \delta) \to (\StExp/{\equiv}, [\delta]_\equiv)\) is a chart homomorphism.
\end{proposition}

\cref{prop:regex mod bisim} has several consequences, including the following two observations, which appear as~\cite[Proposition~5.1]{grabmayer-fokkink-2020} and~\cite[Proposition~2.9]{grabmayer-fokkink-2020}.

\begin{proposition}%
    \label{prop:regex solution pullback}
    Let \(h \colon (X, \delta_X) \to (Y, \delta_Y)\) be a chart homomorphism, and let \(\phi\colon Y \to \StExp\) be a solution.
    Then \(\phi \circ h\) is a solution to \((X, \delta_X)\).
    Furthermore, for any \(r \in \StExp\), the inclusion map \(\langle r \rangle \into \StExp\) is a solution to \(\langle r \rangle\).
\end{proposition}

Grabmayer and Fokkink's completeness proof strategy requires the construction of a distinguished class of charts \(\mathcal C\) satisfying all of the following properties:
\begin{description}
    \item[(Expressivity)] For each \(r \in \StExp\), the chart \(\langle r \rangle\) is in \(\mathcal C\).
    \item[(Closure)]\label{prop:closure} \(\mathcal C\) is \emph{closed under homomorphic images}.
    That is, if there is a surjective chart homomorphism \((X, \delta_X) \to (Y, \delta_Y)\) and \((X, \delta_X) \in \mathcal C\), then \((Y, \delta_Y) \in \mathcal C\).
    \item[(Solvability)] Every chart \((X, \delta) \in \mathcal C\) admits a unique solution.
\end{description}
The restriction of surjectivity in the second property is relatively mild: any chart homomorphism $h: (X, \delta_X) \to (Y, \delta_Y)$ can be restricted to a homomorphism onto its image via standard techniques~\cite{rutten-00}.
If a class satisfying these properties exists, then the completeness of \(\SL^*\) for bisimilarity can be argued as follows.

\begin{proof}[\cref{thm:gf completeness}, completeness direction]
    Let \(r_1,r_2 \in \StExp\) and suppose \(r_1 \bisim r_2\).
    Then there is a chart \((Z, \delta)\) and chart homomorphisms \(h_i \colon \langle r_i\rangle \to (Z, \delta)\) for $i \in \{1, 2\}$ such that \(h_1(r_1) = h_2(r_2)\).
    Let \(z = h_1(r_1) = h_2(r_2)\), and note that \(\langle z\rangle = h_1(\langle r_1\rangle) = h_2(\langle r_2 \rangle)\).
    In particular, \(\langle z\rangle\) is the homomorphic image of \(\langle r_1\rangle\).
    By \textbf{(Expressivity)}, \(\langle r_1\rangle \in \mathcal C\).
    By \textbf{(Closure)}, \(\langle z\rangle \in \mathcal C\), because it is the homomorphic image of \(\langle r_1\rangle\).
    By \textbf{(Solvability)}, \(\langle r_1\rangle\), \(\langle r_2\rangle\), and \(\langle z\rangle\) all admit unique solutions.
    Let \(\phi \colon \langle z \rangle \to \StExp\) be the solution to \(\langle z \rangle\).
    By \cref{prop:regex solution pullback}, \(\phi \circ h_i\) is a solution to $\langle r_i \rangle$ for $i \in \{1,2\}$.
    Since the inclusions \(\langle r_i \rangle \into \StExp\) are also solutions, by uniqueness
    \(
        \SL^* \vdash r_1 = \phi \circ h_1(r_1) = \phi \circ h_2(r_2) = r_2
    \), as desired.
\end{proof}

\paragraph*{Well-layered Charts and Solutions.}
For the rest of the section, we focus on a class \(\mathcal C\) that satisfies the three properties above.
Milner already observed that not every chart admits a solution~\cite{milner-84}.
This stands in sharp contrast with the situation for regular expressions considered \emph{up to language equivalence}, where every automaton can be transformed into an equivalent regular expression by Kleene's theorem.
To address this, Grabmayer and Fokkink proposed \emph{LLEE charts}, which were later refined into \emph{well-layered} charts in~\cite{schmid-rot-silva-21}.

\begin{definition}[{\cite[Definition 4.1]{schmid-rot-silva-21}}]%
    \label{def:well-layered}
    Let \((X, \delta)\) be a chart.
    An \emph{entry/body labelling} of \((X, \delta)\) is a partition of \({\to_\delta} \subseteq X \times X\) into two relations: \emph{loop entry} transitions, denoted \(\eo\), and \emph{body} transitions, denoted \(\bo\).

    We write \(x \diredge y\) and say that \(x\) \emph{loops-around-to} \(y\) if there is a sequence of transitions \(x \eo x_1 \bo \cdots \bo x_n = y\) such that \(x \notin \{x_1, \dots, x_n\}\).

    An entry/body labelling of \((X, \delta)\) is \emph{well-layered} if it satisfies the following additional properties.
    Write \((-)^+\) for transitive closure.
    \begin{enumerate}
        \item We do not have \(x \bo^+ x\) for any \(x \in X\).
        \item For any \(x,y \in X\), if \(x \eo y\), then \(y \bo^+ x\).
        \item The directed graph \((X, \diredge)\) is acyclic.
        \item For any \(x,y \in X\), if \(x \diredge y\), then we do not have \(y \to \checkmark\).
    \end{enumerate}
    A chart is \emph{well-layered} if (1)~it has a well-layered entry-body labelling, and (2)~is \emph{locally finite} in the sense that for any $x \in X$, \(\langle x \rangle\) is finite.
\end{definition}

As we have already seen, \((\StExp, \delta)\) is locally finite.
The equivalence between LLEE charts and well-layered charts is explained in~\cite[Remark 4.1]{schmid-rot-silva-21}.

The loop entry transitions in a well-layered entry/body labelling are to be interpreted as the first transition that enters a program loop.
This is formally captured by the entry/body labelling on \((\StExp, \delta)\) given as follows: loop entry transitions are those that can be derived from the rules
\begin{gather}
    \label{eq:entry/body of regex}
    \infer{r_1 \to \checkmark}{r_1 * r_2 \eo r_1 * r_2}
    \qquad
    \infer{r_1 \to s \quad s \to^+ \checkmark}{r_1 * r_2 \eo s(r_1 * r_2)}
    \qquad
    \infer{r_1 \eo s}{r_1r_2 \eo s r_2}
\end{gather}
Body transitions are all other transitions.
The following result is stated as~\cite[Lemma 4.1]{schmid-rot-silva-21}, but follows directly from~\cite[Proposition 3.7]{grabmayer-fokkink-2020} and the equivalence between LLEE charts and well-layered charts.

\begin{proposition}%
    \label{prop:regex well-layered}
    The entry/body labelling of the chart \((\StExp, \delta)\) given in~\eqref{eq:entry/body of regex} is well-layered.
    It follows that \(\langle r \rangle\) is well-layered for any \(r \in \StExp\).
\end{proposition}

The second statement in \cref{prop:regex well-layered} follows from the first:
it is a direct consequence of \cref{def:well-layered} and the properties of subcharts that if \(\eo,\bo\) is a well-layered entry/body labelling of \((X, \delta)\), then \({\eo}\cap U^2,{\bo}\cap U^2\) is also a well-layered entry/body labelling of the generated subchart \(\langle U \rangle_\delta\).

If we choose the class of well-layered charts for $\mathcal{C}$, then \cref{prop:regex well-layered} is \textbf{(Expressivity)}.
\textbf{(Closure)} is stated below as \cref{thm:well-layered covariety}.
\cref{thm:well-layered covariety} is a mild strengthening of~\cite[Theorem 6.9]{grabmayer-fokkink-2020} that appears as~\cite[Theorem 4.1]{schmid-rot-silva-21}.

\begin{theorem}[Grabmayer-Fokkink]%
    \label{thm:well-layered covariety}
    The class of well-layered charts is \emph{closed under homomorphic images}.
\end{theorem}

\cref{thm:well-layered covariety} is the crux of the completeness proof in~\cite{grabmayer-fokkink-2020}, and took an enormous amount of ingenuity to prove.
We will rely on this result later, in \cref{sec:completeness}, when we prove our general completeness theorem.

Finally, let us discuss \textbf{(Solvability)}, i.e., existence and uniqueness of solutions, which is the last of the pieces needed in the completeness proof for 1-free star expressions.
Grabmayer and Fokkink offer the following inductively defined formula for computing the unique solution to a well-layered chart at each state.

\begin{definition}%
    \label{def:regex canonical solution}
    Given a well-layered entry/body labelling \(\eo,\bo\) of a chart \((X, \delta)\), define the following two quantities: given \(x \in X\), let
    \(
        |x|_{en} = \max\{n \mid \exists y \in Y, x \diredge^n y\}
    \)
    and
    \(
        |x|_{bo} = \max\{n \mid \exists y \in X, x \bo^n y\}
    \).

    We define $\phi_\delta: X \to \StExp$ inductively on \(|x|_{bo}\) as follows:
    \begin{equation}
        \label{eq:regex canonical sol}
        \phi_\delta(x)
        = \Big(\sum_{x \etr{a} x} a + \sum_{\substack{x \etr{a} y\\x \neq y}} a~t_\delta(y, x)\Big)
            * \Big(\sum_{x \tr{a} \checkmark} a + \sum_{x \btr{a} y} a\phi_\delta(y)\Big)
    \end{equation}
    in which we define for each pair of states such that \(x \diredge y\) the term $t_\delta(y, x)$ below, by induction on the lexicographical ordering of $\N \times \N$:
    \begin{equation}
        \label{eq:regex canonical t}
        t_\delta(y, x)
        = \Big(\sum_{y \etr{a} y} a + \sum_{\substack{y \etr{a} z\\y \neq z}} a~t_\delta(z, y)\Big)
            * \Big(\sum_{y \btr{a} x} a + \sum_{\substack{y \btr{a} z\\x \neq z}} a~t_\delta(z, x)\Big)
    \end{equation}
\end{definition}

The following corresponds to~\cite[Proposition 5.5]{grabmayer-fokkink-2020} and~\cite[Proposition 5.8]{grabmayer-fokkink-2020}, and establishes \textbf{(Solvability)} for well-layered charts.

\begin{proposition}%
    \label{prop:regex canonical solution}
    Let \((X, \delta)\) be a well-layered chart with entry/body labeling \(\eo,\bo\).
    Then \(\phi_\delta\) as derived from \(\eo,\bo\) in \cref{{def:regex canonical solution}} is the unique solution to \((X, \delta)\).
    In particular, $\phi_\delta$ does not depend on $\eo, \bo$ up to equivalence.
\end{proposition}


\section{Equational Theories and \(M\)-systems}%
\label{sec:M-systems}

1-free star expressions denote processes that can be composed nondeterministically using $+$, and include the constant $0$ for the process without outgoing transitions.
The axioms involving $+$ and $0$ are precisely those of a join-semilattice with bottom (see \Cref{inst:semilattices} below).
At the same time, the \emph{free} semilattice with bottom generated by a set \(X\) is \(\P(X)\), the finite powerset of \(X\).
This monad also pops up in the transition functions for charts, which are of the form \(X \to \P(\Act \times( \checkmark + X))\).

This is not a coincidence, and in this section we study the connection more formally.
First, we recall \emph{equational theories}, and touch on a well-known connection to \emph{free algebra constructions}.
This leads to an abstracted notion of a chart, parametrized by an equational theory, with its own notion of bisimilarity.
The sections that follow develop these ideas to obtain a parametrized axiomatization.


\paragraph*{Equational Theories.}
A \emph{signature} is a set of \emph{operation symbols} \(S\) paired with a function \(\ar\colon S\to \N\).
The value \(\ar(\sigma)\) is called the \emph{arity} of \(\sigma \in S\).
The set of \emph{\(S\)-terms over \(X\)}, \(S^*X\), is the smallest set of formal expressions containing $X$, and such that if $\sigma \in S$ with $\ar(\sigma) = n$ and $t_1, \dots, t_n \in S^*X$, then $\sigma(t_1, \dots, t_n) \in S^*X$.

Given \(t \in S^*X\) and \(\nu \colon X \to S^*Y\), we write \(t(\nu)\) to denote the term in \(S^*Y\) obtained by replacing each \(x\) in \(t\) with \(\nu(x)\).
We will often write \(t = t(x_1, \dots, x_n)\) for distinct \(x_1, \dots, x_n\) to signal that the variables in \(t\) are among \(x_1, \dots, x_n\), and more compactly \(t(\vec x)\) where \(\vec x = (x_1, \dots, x_n)\).
Given \(t = t(\vec x)\) and \(t_1, \dots, t_n \in S^*Y\), we write \(t(t_1, \dots, t_n)\) for the term \(t(\nu)\) where \(\nu \colon X \to S^*Y\) is such that \(\nu(x_i) = t_i\).

Fix a set of \emph{variables} \(\Var\) and a signature $S$.
Any relation on $S^*\Var$ can be seen as a set of \emph{(formal) equations} over $S$-terms.
Given \(S\)-terms \(t, s\) over \(\Var\) and a set of equations $\mathsf E$, we write \(\mathsf E \vdash t = s\) if \(t = s\) can be derived from the equations in \(\mathsf E\) and the laws of equational logic (reflexivity, symmetry, transitivity, substitution, and congruence).
An \emph{equational theory} for $S$ is a set of equations that is closed under the inference rules of equational logic.
A set of equations $\mathsf E$ is an \emph{axiomatization} of \(\mathsf T\) if \(\mathsf T\) is the smallest equational theory containing \(\mathsf E\).
In the future, we abuse terminology and simply refer to \(\mathsf E\) as an equational theory when we are in fact referring to the equational theory it axiomatizes.


An \emph{\(S\)-algebra} is a pair \((X, \rho)\) consisting of a set \(X\) and for each \(\sigma \in S\) an operation \(\sigma^\rho \colon X^{\ar(\sigma)} \to X\).
An \(S\)-algebra homomorphism \(h \colon (X, \rho_X) \to (Y, \rho_Y)\) is a function \(h \colon X \to Y\) such that \(h(\sigma^{\rho_X}(x_1, \dots, x_n)) = \sigma^{\rho_Y}(h(x_1), \dots, h(x_n))\).

Given an \(S\)-algebra \((X, \rho)\) and \(t \in S^*X\), we can evaluate \(t\) to $t^\rho \in X$ by setting \(x^\rho = x\) and \(\sigma(t_1, \dots, t_n)^\rho = \sigma^\rho(t_1^\rho, \dots, t_n^\rho)\).
An \(S\)-algebra \((X, \rho)\) \emph{satisfies} a set of equations \(E\), written \((X, \rho) \models E\), if \((t_1, t_2) \in E\) implies \(t_1^\rho = t_2^\rho\).

We can give an equivalent description of \(S\)-algebras using some categorical language.
If we write \(\Set\) for the category of sets and functions, we can form the signature functor \(\Sigma_S = \bigsqcup_{\sigma \in S} \{\sigma\}\times\Id^{\ar(\sigma)}\) out of a signature \(S\), where \(\Id\) is the identity functor on \(\Set\).
Then an \(S\)-algebra is the same data as a function \(\rho \colon \Sigma_S X \to X\).
We usually abuse notation and simply write \(S\) instead of \(\Sigma_S\).

\begin{definition}%
    \label{def:free algebra}
    Let \(\T\) be an equational theory for the signature \(S\).
    A \emph{free-algebra construction} for \(\T\) is a triple \((M, \eta, \rho)\) consisting of an endofunctor \(M\) on \(\Set\) and natural transformations \(\eta \colon \Id \Rightarrow M\) and \(\rho \colon SM \Rightarrow M\) such that \((MX, \rho_X) \models \T\) for all \(X\), satisfying the following: given an \(S\)-algebra \((Y, \mu)\) that satisfies \(\T\), and given \(f \colon X \to Y\), there is a unique \(S\)-algebra homomorphism \(f^\# \colon (MX, \rho_X) \to (Y, \mu)\) such that \(f^\# \circ \eta = f\), called the \emph{Kleisli extension} of \(f\).
\end{definition}

Any two free-algebra constructions for the same equational theory are isomorphic, in the sense that there exist natural isomorphisms between their functors that interacts well with the other structure.
Thus, we often refer to a free-algebra construction for \(\T\) as \emph{the} free-algebra construction for \(\T\).

\begin{remark}
    Free-algebra constructions for equational theories as we have described them have a close relationship with \emph{finitary monads} on \(\Set\).
    In particular, the free-algebra constructions for an equational theory are in one-to-one correspondence with monads \emph{presented by} the theory~\cite{bonchi-sokolova-vignudelli-21}, and it is known that a monad is finitary if and only if it has an equational presentation~\cite{adamek-rosicky-94}.
\end{remark}


We will use these instantiations of \cref{def:free algebra} in the rest of the paper.

\begin{instantiation}%
    \label{inst:semilattices}
    The \emph{theory of semilattices (with bottom)} is the equational theory $\SL$ for the signature \(S = \{+,0\}\) that axiomatized by the following equations:
    \begin{mathpar}
        x + 0 = x
        \and
        x + x = x
        \and
        x + y = y + x
        \and
        x + (y + z) = (x + y) + z
    \end{mathpar}
    If we define \(\eta_X(x) = \{x\}\) and \(\rho_X \colon S\P X \to \P X \) by \(V_1 +^{\rho_X} V_2 = V_1 \cup V_1\) and \(0^{\rho_X} = \emptyset\), then \((\P, \{-\}, \rho)\) is a free-algebra construction for $\SL$ (keep in mind that we use \(\P\) for the \emph{finite} powerset).
    The Kleisli extension of \(f\colon X \to Y\) with \((Y, +^\mu, 0^\mu)\) a semilattice is \(f^\#\colon \P X \to Y\) defined by \(f^\#(\{ x_1, x_2, \dots, x_n \}) = f(x_1) +^\mu f(x_2) +^\mu \cdots +^\mu f(x_n)\), with the empty sum being $0^\mu$.
    The theory of semilattices describes the branching type of 1-free star expressions.
\end{instantiation}

\begin{instantiation}%
    \label{inst:guarded alg}
    Let \(T\) be a finite set of \emph{primitive tests} and let \(\mathit{BA}\) be the free Boolean algebra on \(T\), i.e., the set of Boolean expressions generated from \(T\) using $\bot$, $\wedge$ and $\neg$, modulo the axioms of Boolean algebra.
    The signature $S$ of \emph{guarded algebra} has a constant \(0\) and a binary operation \(+_b\) for each \(b \in \mathit{BA}\).
    Its theory $\GA$ is axiomatized by the following equations for all $b, c \in \mathit{BA}$:
    \begin{mathpar}
        x +_b x = x
        \and
        x +_b y = y +_{\neg b} x
        \and
        (x +_b y) +_c z = x +_{b\wedge c} (y +_c z)
    \end{mathpar}
    Let \(\At\) be the set of atomic elements of the Boolean algebra \(\mathit{BA}\) and note that this set is finite.
    Then the free algebra construction for \(\GA\) is the triple \((\R, \eta, \rho)\) where \(\R\) is the \emph{reader-with-exception functor} \(\R X = (\bot + X)^{\At}\), and for each \(\alpha \in \At\), \(\eta_X(x)(\alpha) = x\), \((\theta_1 +_b^{\rho_X} \theta_2)(\alpha) = \mathbf{if}~\alpha \le b~\mathbf{then}~\theta_1(\alpha)~\mathbf{else}~\theta_2(\alpha)\), and \(0^{\rho_X}(\alpha) = \bot\).
    It is helpful to think of \(x +_b y\) as notation for \(\mathbf{if}~b~\mathbf{then}~x~\mathbf{else}~y\).
    The theory of guarded algebra captures the branching of skip-free GKAT programs~\cite{kappe-schmid-silva-2023}.
\end{instantiation}

\begin{instantiation}%
    \label{inst:convex alg}
    The signature $S$ of \emph{(positive) convex algebra} consists of one constant symbol \(0\) and a binary operation \(\oplus_p\) for each \(p \in [0,1]\).
    Its theory $\CA$ is axiomatized by the following equations for all $p, q \in [0,1]$:
    \begin{gather*}
        x \oplus_1 y = x
        \qquad
        x \oplus_p x = x
        \qquad
        x \oplus_p y = y \oplus_{(1 - p)} x
        \\
        (x \oplus_p y) \oplus_q z = x \oplus_{pq} (y \oplus_{(q(1-p)/(1 - pq))}) \quad \text{(if $pq < 1$)}
    \end{gather*}
    Let \(\D X\) denote the set of finitely supported probability distributions on \(X\), and for each \(x \in X\) define the \emph{Dirac delta} distribution \(\delta_x(y) = \textbf{if}~x = y~\textbf{then}~1~\textbf{else}~0\).
    Then the triple \((\D(\bot + (-)), \eta, \rho)\) is a free algebra construction for \(\CA\)~\cite{swirszcz-74}, where \(\eta_X(x) = \delta_x\), \(\theta_1 \oplus_p^{\rho_X} \theta_2 = p\theta_1 + (1-p)\theta_2\), and \(0^{\rho_X} = \delta_\bot\).
    The theory of convex algebra captures the branching of probabilistic processes~\cite{stark-smolka-00}.
\end{instantiation}

\paragraph{\(M\)-systems.}
Fix an equational theory \(\T\) for the signature \(S\) and a free-algebra construction \((M, \eta, \rho)\) for \(\T\).
At the core of a free-algebra construction \((M, \eta, \rho)\) for \(\T\) is its endofunctor \(M\) on \(\Set\).
By replacing the finite powerset functor \(\P\) in the transition function type for charts \(X \to \P(\Act \times (\checkmark + X))\) with \(M\), we obtain a transition function type for processes whose branching is captured by \(\T\).

\begin{definition}%
    \label{def:M-system}
    Let \(B_M\) be the endofunctor \(M(\Act\times(\checkmark + (-)))\) on \(\Set\).
    Then an \emph{\(M\)-system} is a pair \((X, \beta)\) consisting of a set of \emph{states} \(X\) and a \emph{transition function} \(\beta\colon X \to B_M X\).
    A \emph{homomorphism} of \(M\)-systems \(h\colon (X, \beta_X) \to (Y, \beta_Y)\) is a function \(h \colon X \to Y\) such that \(B_M(h) \circ \beta_X = \beta_Y \circ h\).
\end{definition}

\(M\)-systems are precisely \(B_M\)-coalgebras, and homomorphisms of \(M\)-systems are precisely $B_M$-coalgebra homomorphisms.
This unlocks many useful results from coalgebra~\cite{rutten-00}.
In particular, \(M\)-systems and their homomorphisms form a category, which we will call \(\Coalg(B_M)\).
The theory of coalgebras also prescribes a general notion of bisimilarity, which instantiates to \cref{def:chart bisimilarity} via \Cref{lem:characterize chart bisim}.


\begin{definition}%
    \label{def:bisimilarity}
    Let \((X, \beta_X)\) and \((Y, \beta_Y)\) be \(M\)-systems.
    We call \(x \in X\) and \(y \in Y\) \emph{bisimilar} (written \(x \bisim y\)) if there is an \(M\)-system \((Z, \beta_Z)\) and homomorphisms \(h \colon (X, \beta_X) \to (Z, \beta_Z)\) and \(k \colon (Y, \beta_Y) \to (Z, \beta_Z)\) such that \(h(x) = k(y)\).
\end{definition}

A standard argument tells us that bisimilarity is an equivalence relation~\cite{jacobs-16}.

\section{Skip-free Star Expressions}
Grabmayer and Fokkink introduced 1-free star expressions as a specification language for processes with non-deterministic branching behaviour captured by charts.
As we saw, charts coincide with \(\P\)-systems.
Moreover, the algebraic signature of the theory of semilattices --- for which $\P$ provides the free algebra construction --- suggests a syntax for behaviour expressed by well-layered charts, as well some axioms.
In this section, we expand on these ideas, using equational theories and free algebra constructions to develop a syntax for $M$-systems, as well as a candidate set of axioms for equivalence expressed in this syntax.

For the remainder of this section, we fix an equational theory \(\T\) for the signature \(S\), which admits a free algebra construction $(M, \eta, \rho)$.

\begin{definition}%
    \label{def:skip-free star exp}
    The \emph{skip-free unified-star expressions} \(\Exp\) are generated by
    \[
        \Exp \ni e_i ::= \sigma(e_1, \dots, e_n) \mid a \in \Act \mid e_1\cdot e_2 \mid e_1^{(s)}e_2
    \]
    In the above, \(\sigma \in S\) with \(\ar(\sigma) = n\) and \(s = s(u, v)\) is an \(S\)-term in two variables.
\end{definition}

The small-step semantics of skip-free star expressions is captured by giving \(\Exp\) the structure of an \(M\)-system.
This automatically equips these expressions with a notion of equivalence, namely bisimilarity as states in this system.
We call the \(M\)-system \((\Exp, \gamma)\), defined recursively in \cref{fig:synt M-system}, the \emph{syntactic \(M\)-system}.

Note that we use the following shorthands from now on: \((\vec a, \vec e)\) denotes the list of tuples \((a_1, x_1), \dots, (a_n, x_n)\); \(\vec b \vec e\) denotes the list \(b_1e_1, \dots, b_n e_n\); \(\vec e f\) denotes the list \(e_1f,\dots,e_n f\); and generally, where \(h\) is a function defined on the set \(\{x_1, \dots, x_n\}\), \(h(\vec x)\) denotes the list \(h(x_1), \dots, h(x_n)\).

\begin{figure}[t]
    \begin{gather*}
        \gamma(a) = \eta(a, \checkmark)
        \qquad
        \gamma(\sigma(\vec e)) = \sigma^\rho(\gamma(\vec e))
        \qquad
        \gamma(e_1e_2) = t^\rho((\vec b, e_2), (\vec a, \vec fe_2))
        \\
        \begin{aligned}
        \gamma(e_1^{(\textcolor{red}{s})}e_2)
        &= \textcolor{red}{s^\rho\Big(} t^\rho((\vec b, e_1^{(s)}e_2), (\vec a, \vec f(e_1^{(s)}e_2))) \textcolor{red}{,} ~  \gamma(e_2) \textcolor{red}{\Big)} 
        \end{aligned}
    \end{gather*}
    \vspace*{-2em}
    \caption{
        The definition of \((\Exp, \gamma)\) w.r.t.~the free-algebra construction \((M, \eta, \rho)\).
        Above, \(e_i \in \Exp\), \(a \in \Act\), and \(s = s(u,v) \in S^*\Var\).
        In the formulas for \(e_1e_2\) and \(e_1^{(s)}e_2\) above, \(\gamma(e_1) = t^\rho = t^\rho((\vec b, \checkmark), (\vec a, \vec f))\).
        In the last formula, we have used \textcolor{red}{red} to indicate the portion of the formula that corresponds to the \(s\) in \(e_1^{(s)}e_2\).
    }%
    \label{fig:synt M-system}
\end{figure}

Intuitively, the skip-free star expressions of the form \(a \in \Act\) and \(e_1 \cdot e_2\) have the usual interpretation: \(a\) is the process that emits \(a\) and accepts, and \(e_1 \cdot e_2\) is the sequential composition of \(e_1\) and \(e_2\).
As before, we usually write \(e_1e_2\) instead of \(e_1\cdot e_2\).
The expression \(\sigma(e_1, \dots, e_n)\) denotes the process that branches into \(e_1, \dots, e_n\) and whose outgoing transitions carry the structure defined by \(\sigma\).

\begin{example}%
\label{example:nondet-simple}
Let $\T = \SL$ be the theory of join-semilattices with bottom discussed in \Cref{inst:semilattices}.
The free-algebra construction $(M, \eta, \rho)$ for \(\T\) is \((\P, \{{-}\}, \rho)\), and the behaviour of $(\Exp, \gamma)$ corresponds closely to that of $(\StExp, \delta)$ (cf.\ \Cref{fig:syntactic chart}).

For instance, $\gamma(a) = \eta(a, \checkmark) = \{ (a, \checkmark) \}$, and similarly $\gamma(b) = \{ (b, \checkmark) \}$.
This matches the transitions of $a, b \in \StExp$ as prescribed by $\delta$.
Furthermore, $\gamma(a + b) = \gamma(a) +^\rho \gamma(b) = \gamma(a) \cup \gamma(b) = \{ (a, \checkmark), (b, \checkmark) \}$, which matches $\delta(a+b)$.

If we look at $(a + b)c \in \Exp$, then \Cref{fig:synt M-system} tells us that because $\gamma(a + b) = \eta(a, \checkmark) +^\rho \eta(b, \checkmark)$, we have $\gamma((a + b)c) = \eta(a, c) +^\rho \eta(b, c) = \{ (a, c), (b, c) \}$; this also corresponds to the transitions exiting $(a + b)c \in \StExp$ as specified by $\delta$.
\end{example}

The process denoted \(e_1^{(s)}e_2\), given by the \emph{\(s\)-star} of \(e_1\) and \(e_2\), is a bit more complicated: it represents the process that loops on the branches of \(e_1\) wherever the variable \(u\) appears in \(s(u, v)\), and otherwise moves on to the branches of \(e_2\) wherever the variable \(v\) appears.
In the future, if \(s(u, v) = \sigma(u, v)\) for some binary operation \(\sigma \in S\), we will write \(e_1^{(\sigma)}e_2\) in place of \(e_1^{(\sigma(u, v))}e_2\).

\begin{example}%
\label{example:nondet-star}
Let $M$, $\T$ and the free-algebra construction be as in the previous example.
We can recover the behaviour of the binary star operator from $1$-free regular expressions as a particular instance of skip-free unified-star expressions.

For example, if we  look at $(a + b) * c$, then \Cref{fig:syntactic chart} tells us that
\[
    \gamma(a * b) = \{ (a, (a+b)*c), (b, (a+b)*c), (c, \checkmark) \}
\]
At the same time, if we consider $a^{(+)}b$, then \Cref{fig:synt M-system} tells us that because $\gamma(a + b) = \eta(a, \checkmark) +^\rho \eta(b, \checkmark)$, we have that
\begin{align*}
    \gamma((a+b)^{(+)}c)
        &= (\eta(a, (a+b)^{(+)}c) +^\rho \eta(b, (a+b)^{(+)}c)) +^\rho \eta(c, \checkmark) \\
        &= \{ (a, (a+b)^{(+)}c), (b, (a+b)^{(+)}c), (c, \checkmark) \}
\end{align*}
In fact, the above can be used to show that $(a+b)*c$ as a state in $(\StExp, \delta)$ is bisimilar to $(a+b)^{(+)}c$ as a state in $(\Exp, \gamma)$.

Up to equivalence, there are three more choices for $s = s(u, v)$ in $e_1^{(s)}e_2$, namely the variables $u$ and $v$, and the constant $0$.
The behaviours obtained by these expressions can still be modelled by $1$-free regular expressions: $a^{(u)}b$ corresponds to $a * 0$, $a^{(v)}b$ corresponds to $0 * a$, and $a^{(0)}b$ is simply $0$.
\end{example}

Given \(M\)-systems \((U, \beta_U)\) and \((X, \beta)\), \((U, \beta_U)\) is a \emph{subsystem} of \((X, \beta)\) if \(U \subseteq X\) and the inclusion map \((U, \beta_U) \into (X, \beta)\) is a homomorphism of \(M\)-systems.
The syntactic $M$-system has a finite subsystem for each expression.

\begin{restatable}{proposition}{propLocallyFinite}%
    \label{prop:Exp locally finite}
    For each \(e \in \Exp\), there is a smallest finite subsystem \(\langle e \rangle \) of \((\Exp, \gamma)\) containing \(e\), called the subsystem \emph{generated by \(e\)}.
\end{restatable}

\begin{figure}[t]
    \centering
    \begin{gather*}
        \begin{gathered}
            (\T) \quad \infer{\T \vdash t(\vec x) = s(\vec x)}{t(\vec e) = s(\vec e)}
        \end{gathered}
        \qquad
        \begin{aligned}
            (\mathrm{A}) && e(fg) &= (ef)g \\
            (\mathrm{D}) && t(\vec e)f &= t(\vec ef) \\
            (\mathrm{U}) && e^{(s)}f &= s(e(e^{(s)}f), f)
        \end{aligned}
        \qquad
        \begin{gathered}
            (\mathrm{RSP}) \quad \infer{g = s(eg, f)}{g = e^{(s)}f}
        \end{gathered}
    \end{gather*}
    \vspace*{-2em}
    \caption{
        The theory \(\T^*\) that axiomatizes bisimilarity of star expressions.
        Above, \(e_1,\dots,e_n  \in \Exp\), \(e,f,g \in \Exp\), \(t = t(\vec v) \in S^*\Var\), and \(s(u, v) \in S^*\Var\) is a binary term with free variables \(u, v\).
        Recall that the notation \(\vec e f\) means \(e_1f, \dots, e_n f\).
    }%
    \label{fig:axioms}
\end{figure}

We now present a set of axioms that aims to capture bisimilarity in \((\Exp, \gamma)\).

\begin{definition}%
    \label{def:axioms}
    The \emph{skip-free unified-star theory} \(\T^*\) consists of the axioms in \cref{fig:axioms} and the laws of equational logic.
    We write \(\T^* \vdash e_1 = e_2\) if \(e_1 = e_2\) is derivable from \(\T^*\) and say that \(e_1\) and \(e_2\) are \emph{provably equivalent}.
\end{definition}

Intuitively, the first inference rule in \cref{fig:axioms} says that two terms that are equivalent up to \(\T\) are equivalent as branching structures.
The rules (A) and (D) express standard properties of sequential composition: associativity and right distributivity over branches.
The rules (U) and (RSP) state that \(e_1^{(s)}e_2\) is the unique process $z$ that satisfies the recursion equation \(z = s(e_1z, e_2)\).

\begin{restatable}[Soundness]{theorem}{thmSoundness}%
    \label{thm:soundness}
    Let \(e_1, e_2 \in \Exp\).
    If \(\T^* \vdash e_1 = e_2\), then \(e_1 \bisim e_2\).
\end{restatable}

Even working with an arbitrary equational theory, a number of interesting equivalences can be proven using the axioms of \(\T^*\).
For example, \(\T^* \vdash e_1^{(s)}(e_2e_3) = (e_1^{(s)}e_2)e_3\) is a consequence of (A), (D), (U), and (RSP).
Also, if \(\T \vdash s_1 = s_2\), then \(\T^* \vdash e_1^{(s_1)}e_2 = e_1^{(s_2)}e_2\), which is a consequence of (\(\T\)), (U), and (RSP).


In \Cref{example:nondet-simple,example:nondet-star}, we have already seen that for $\SL$, we recover a system that corresponds to $1$-free star expressions.
For $\GA$ and $\CA$, we are in a similar situation: every \(s\)-star operation \(e_1^{(s)}e_2\) is equivalent (up to the unified-star axioms) to a binary star, i.e., either \(e_1^{(+_b)}e_2\) or \(e_1^{(\oplus_p)}e_2\) respectively.
These correspond precisely to the binary stars in the syntax for skip-free GKAT~\cite{kappe-schmid-silva-2023} and in the probabilistic regular expressions in~\cite{rozowskiS24}.
Thus, our parametrized class of skip-free process algebras does indeed capture both skip-free GKAT and the algebra of skip-free probabilistic regular expressions (modulo bisimilarity).

\section{Completeness}%
\label{sec:completeness}

We are now ready to start proving the completeness of \(\T^*\) w.r.t.\ bisimilarity in \((\Exp, \gamma)\), at least for a large class of equational theories \(\T\).
We follow the same steps as Grabmayer and Fokkink, adapting and generalizing each step to other equational theories.
Specifically, we will construct a class \(\mathcal C\) of \(M\)-systems that satisfies analogues of \textbf{(Expressivity)}, \textbf{(Closure)} and \textbf{(Solvability)}.

Note that we are not able to prove completeness of \(\T^*\) for arbitrary \(\T\).
We will discuss which equational theories we have to restrict our attention to later in this section.
We delay the restriction for now because the first few results presented below are true for arbitrary equational theories.

\paragraph*{Solutions to \(M\)-systems.}
The completeness proof to follow casts \(M\)-systems as systems of equations of a certain form.
For a given \(M\)-system \((X, \beta)\), we associate each state \(x \in X\) with an equation \(x = t(\vec b, \vec a \vec x)\) with unknowns \(x_1, \dots, x_n\), where \(t \in S^*X\) and \(\beta(x) = t^\rho((\vec b, \checkmark), (\vec a, \vec x))\).

\begin{definition}%
    \label{def:solution}
    A \emph{solution} to an \(M\)-system \((X, \beta)\) is a map \(\phi \colon X \to \Exp\) such that for any \(x \in X\) with \(\beta(x) = t^\rho((\vec b, \checkmark), (\vec a, \vec x))\), \(\T^* \vdash \phi(x) = t(\vec b, \vec a\phi(\vec x))\).
    Two solutions \(\phi,\psi\) are \emph{equivalent} if \(\T^*\vdash \phi(x) = \psi(x)\) for all \(x \in X\).
    The \(M\)-system \((X, \beta)\) \emph{admits a unique solution} if it has exactly one solution up to equivalence.
\end{definition}

Note that the specific choice of \(t\) in \cref{def:solution} does not change the space of solutions to \((X, \beta)\).
The following result is analogous to \cref{prop:regex mod bisim}.

\begin{restatable}{proposition}{propSolutionsAreHomoms}%
    \label{prop:solutions are homoms}
    There is a unique \(M\)-system structure \((\Exp/{\equiv}, [\gamma]_\equiv)\) such that the quotient map \([-]_\equiv \colon (\Exp, \gamma) \to (\Exp/{\equiv}, [\gamma]_\equiv)\) is a homomorphism of \(M\)-systems.
    Moreover, for any \(M\)-system \((X, \beta)\), a map \(\phi \colon X \to \Exp\) is a solution if and only if \([-]_\equiv \circ \phi \colon (X, \beta) \to (\Exp/{\equiv}, [\gamma]_\equiv)\) is a homomorphism of \(M\)-systems.
\end{restatable}

As before, \cref{prop:solutions are homoms} has the following immediate consequences.

\begin{restatable}{proposition}{propSolutionPullbacks}%
    \label{prop:solution pullbacks}
    Let \(h \colon (X, \beta_X) \to (Y, \beta_Y)\) be a homomorphism of \(M\)-systems, and let \(\phi \colon Y \to \Exp\) be a solution to \((Y, \beta_Y)\).
    Then \(\phi \circ h\) is a solution to \((X, \beta_X)\).
    Furthermore, for any \(e \in \Exp\), the inclusion map \(\langle e\rangle \into \Exp\) is a solution to \(\langle e\rangle\).
\end{restatable}


We would now like to reuse \cref{thm:well-layered covariety}, which says that well-layered charts are closed under homomorphic images, and \cref{def:regex canonical solution}, which computes the canonical solution to a well-layered chart.
To do this, we need to restrict $\T$.

\paragraph*{Support.}
We start by generalizing the notion of well-layeredness.
Here, the idea is that we need a way to talk about how the states of an $M$-system are connected.
This is encapsulated by the notion of \emph{support}, defined below.

\begin{definition}%
    \label{def:supported}
    A \emph{support} for \(\T\) is a natural transformation \(\supp\colon M \Rightarrow \P\) s.t.\ (1)~\(\supp \circ \eta = \{-\}\) and (2)~for $\sigma \in S$ and $t_1, \dots, t_n \in S^*X$, $\supp(\sigma^{\rho}(t_1^{\rho}, \dots, t_n^{\rho})) \subseteq \supp(t_1^{\rho}) \cup \dots \cup \supp(t_n^{\rho})$.
    \(\T\) is \emph{supported} if a support for \(\T\) exists.
\end{definition}

Intuitively, an equational theory is supported if every term has a well-defined set of ``essential variables'' up to provable equivalence.
For example, in the theory of semilattices \(\SL\), the set of essential variables of a term \(x_1 + \cdots + x_n\) is precisely \(\{x_1, \dots, x_n\}\).
I.e., the support is the identity transformation on $\P$.
For guarded algebra ($\GA$), the support takes a function \(\theta \colon \At \to \bot + X\) to its image without \(\bot\), \(\supp(\theta) = \theta(\At)\setminus\{\bot\}\).
For convex algebra, the support of \(\theta \in \D(\bot + X)\) is its support as a subprobability distribution, \(\supp(\theta) = \{ x \in X \mid \theta(x) > 0\}\).

Not every equational theory is supported.
For a trivial nonexample, the theory axiomatized by \(\mathsf E = \{(u, v)\}\) identifies all terms and has the constant functor \(M = \{\star\}\) and \(\eta_X(x) = \star\) in its free-algebra construction.
The only natural transformation \(\lambda \colon M \Rightarrow \P\) maps \(\star\) to \(\emptyset\), which does not satisfy \(\lambda \circ \eta = \{-\}\).

\begin{remark}
    The existence of a support does not depend on the choice of \((M, \eta, \rho)\), since all free-algebra constructions for \(\T\) are isomorphic.
    So, if one \((M,\eta,\rho)\) has a support, then all do.
    However, more than one support may exist.
\end{remark}

In an \(M\)-system corresponding to a supported equational theory, branching is essentially given by transitions, in the sense of charts, with extra structure.
The \emph{underlying chart} of an \(M\)-system is the chart obtained by forgetting this structure.
This extends the notion of well-layered charts to \(M\)-systems.

\begin{definition}%
    \label{def:underlying chart}
    Let \(\supp\) be a support for \(\T\).
    The \emph{underlying chart} of an \(M\)-system \((X, \beta)\) is the chart \(\supp_*(X, \beta) = (X, \supp_X \circ \beta)\).
    An \(M\)-system is called \emph{well-layered} if its underlying chart is well-layered.
\end{definition}

For the remainder of this section, we shall assume that $\T$ is supported.
In an \(M\)-system \((X, \beta)\), write \(x \tr{a}_\beta \xi\) if \(x \tr{a}_\beta \xi\) in its underlying chart, and \(x \to_\beta y\) if \(x \tr{a}_\beta y\) for some $a$.
By \cref{def:underlying chart}, a well-layered \(M\)-system $(X, \beta)$ admits an entry/body labelling $\eo$, $\bo$ of \(\supp_*(X, \beta)\) that satisfies \cref{def:well-layered}.
We can then write \(x \diredge y\) for states \(x,y \in X\) if \(x \diredge y\) in this entry/body labelling, and furthermore define \(|{x}|_{en}\) and \(|{x}|_{bo}\) as in \cref{def:regex canonical solution} for $x \in X$.

With this candidate class of $M$-systems, we can recover \textbf{(Expressivity)}.

\begin{restatable}{proposition}{propExpWellLayered}%
    \label{prop:Exp well-layered}
    Let \(\supp\) be a support for \(\T\).
    Then \((\Exp, \gamma)\) is well-layered.
    Consequently, for any \(e \in \Exp\), the \(M\)-system \(\langle e \rangle\) generated by \(e\) is well-layered.
\end{restatable}

Recall that $M$-systems are just $B_M$-coalgebras.
The transformation of \(M\)-systems, being a natural transformation, defines a functor \(\supp_* \colon \Coalg(B_M) \to \Coalg(B_\P)\) that has some nice properties~\cite{rutten-00}.
The following is true for general \(F\)-coalgebras for an endofunctor on \(\Set\), and so we state it in full generality.

\begin{restatable}{lemma}{lemInverseImage}%
    \label{lem:inverse image}
    Let \(F\) and \(G\) be endofunctors on \(\Set\) and let \(\lambda \colon F \Rightarrow G\) be a natural transformation.
    Define \(\lambda_* \colon \Coalg(F) \to \Coalg(G)\) to be the functor with \(\lambda_*(X, \beta) = (X, \lambda_X \circ \beta)\) and \(\lambda_*(h) = h\) for any coalgebra homomorphism \(h\).
    Let \(\mathcal C\) be a class of \(G\)-coalgebras that is closed under homomorphic images.
    Then \(\lambda^{-1}\mathcal C = \{(X, \beta) \mid \lambda_*(X, \beta) \in \mathcal C\}\) is also closed under homomorphic images.
\end{restatable}

In our situation, \(F = B_M\), \(G = B_\P\), and \(\lambda = \supp_*\).
As we have defined it, the class of well-layered \(M\)-systems is precisely the inverse image of \(\supp_*\).
From \cref{thm:well-layered covariety,lem:inverse image}, we immediately obtain a version of \textbf{(Closure)}:

\begin{restatable}{theorem}{thmWellLayeredMSystemsCovariety}%
    \label{thm:well-layered M-system covariety}
    Well-layered \(M\)-systems are closed under homomorphic images.
\end{restatable}

\paragraph*{Malleability.}
The point of well-layered systems is that we can solve them uniquely.
To this end, we want to replay the strategy from \cref{def:regex canonical solution}.
The following notion helps to do that, by letting us isolate variables into a subterm.

\begin{definition}%
    \label{def:malleable}
    We say that \(\T\) is \emph{malleable} if for any set \(X\), any partition \(U + V = X\), and any term \(t \in S^*X\), there are terms \(t_1 \in S^*U\), \(t_2 \in S^*V\), and a term \(s = s(u, v) \in S^*\Var\) such that \(\T \vdash t = s(t_1, t_2)\).
\end{definition}

\begin{example}
In the case of $\GA$, if $t = x +_b (y +_c z)$ and $U = \{x, y\}$ while $V = \{ z\}$, then we can choose $s = u +_{b \vee c} v$, $t_1 = x +_b y$ and $t_2 = z$ to find that $\T \vdash t = s(t_1, t_2)$.
More generally, due to the associativity and commutativity properties of \(\SL\), \(\GA\), and \(\CA\), all three of these equational theories are malleable.\footnote{
        These are not necessary conditions for malleability, though. 
        In \cref{sec:examples}, we will see an example of a malleable equational theory that does not enjoy associativity.
}
\end{example}

\begin{remark}
Not all equational theories are malleable.
For a trivial example, consider the theory \(\T = \emptyset\) for a signature with two binary operations \(\star,\bullet\).
Clearly, the term \(x \star (y \bullet z)\) is not of the form \(s(t_1(x, y), t_2(z))\) for any terms \(t_1, t_2, s\).
In \cref{sec:examples}, we will also see an example of a richer equational theory, which captures branching that mixes nondeterminism and probability, that is not malleable.
\end{remark}

For the rest of this section, we assume that $\T$ is malleable.
We can now use this to recover the solution strategy for well-layered charts in \Cref{def:regex canonical solution}.

\begin{definition}%
    \label{def:system canonical solution}
    Let \(\eo,\bo\) be a well-layered entry/body labelling of \((X, \beta)\).
    We define the \emph{canonical solution} to \((X, \beta)\) \emph{given by \(\eo,\bo\)} as follows.
    Let
    \begin{gather*}
        \beta(x) = \textcolor{red}{s^\rho\Big(} t_1^\rho((\vec a, x), (\vec b, \vec x))\textcolor{red}{,}~ t_2^\rho((\vec c, \vec y) , (\vec d, \checkmark))\textcolor{red}{\Big)} 
    \end{gather*}
    where $\vec{x}$ is a vector such that \(x \neq x_i\) and \(x \eo x_i\) for each \(i\), and $\vec{y}$ is a vector such that \(x \bo y_j\) for each \(j\).
    By induction on \(|x|_{bo} \in \mathbb N\), we define
    \begin{equation*}
        \label{eq:star canonical sol}
        \phi_\beta(x)
        = \Big(
                t_1(\vec a, b_1\tau_\beta(x_1, x), \dots, b_n\tau_\beta(x_n, x))
        \Big)^{(s)}\Big(
                t_2(c_1\phi_\beta(y_1), \dots, c_m\phi(y_m), \vec d)
        \Big)
    \end{equation*}
    where, by induction on \((|x|_{en}, |y|_{bo})\) in the lexicographical ordering of \(\N \times \N\), for each pair of states such that \(x \diredge y\) we define \(\tau_\beta(y, x)\) as follows.
    First, let
    \begin{gather*}
        \beta(y) = \textcolor{red}{s^\rho\Big(} t_1^\rho((\vec a, y), (\vec b, \vec x))\textcolor{red}{,}~ t_2^\rho((\vec c, x) , (\vec d, \vec y))\textcolor{red}{\Big)} 
    \end{gather*}
    where $\vec{x}$ is a vector such that \(y \neq x_i\) and \(y \eo x_i\) for each \(i\), and $\vec{y}$ is a vector such that \(x \neq y_k\) and \(y \bo y_k\) for each \(k\).
    Then
    {\footnotesize\[
        \tau_\beta(y, x)
        = \Big(
                t_1(\vec a, b_1\tau_\beta(x_1, y), \dots, b_n\tau_\beta(x_n, y))
        \Big)^{(s)}\Big(
                t_2(\vec c, d_1\tau_\beta(y_1, x), \dots, d_m\tau_\beta(y_m, x))
        \Big)
    \]}
\end{definition}

In the above, we assumed that \(\beta(x)\) and \(\beta(y)\) were in a specific form.
This is where malleability comes in: we partitioned the support of \(\beta(x)\) into pairs that correspond to self loops \(x \eo x\) or loop entry transitions \(x \eo x_i\), and those that come from body transitions \(x \bo y_j\) or accepting transitions \(x \to \checkmark\).
Malleability assures us that we can write $\beta(x)$ as described, and similarly for $\beta(y)$.

Unravelling the definitions in the case of \(\SL^*\), one obtains the canonical solution formula in \cref{def:regex canonical solution}, which appeared in~\cite{grabmayer-fokkink-2020}.
In this case also, \(\phi_\beta\) is the unique solution to a well-layered \(M\)-system, so we recover \textbf{(Solvability)}.

\begin{restatable}{proposition}{propCanonicalSolutionUnique}%
    \label{prop:canonical solution unique}
    Let \((X, \beta)\) be a well-layered \(M\)-system, with entry/body labeling \(\eo,\bo\).
    Then the canonical solution \(\phi_\beta\) given by \(\eo,\bo\) is the unique solution to \((X, \beta)\).
    In particular, up to \(\T^*\), \(\phi_\beta\) does not depend on \(\eo,\bo\).
\end{restatable}

\paragraph{Completeness.}
Following the same steps in the first completeness proof we saw (of \cref{thm:gf completeness}) with \(\mathcal C\) the class of well-layered \(M\)-systems (and replacing the word ``chart'' with ``\(M\)-system'' everywhere), we can apply \cref{prop:regex well-layered,thm:well-layered M-system covariety,prop:canonical solution unique} to obtain the main result of the paper.

\begin{restatable}[Completeness]{theorem}{thmCompleteness}%
    \label{thm:completeness}
    Let \(\T\) be a supported malleable theory for the signature \(S\).
    Given \(e_1, e_2 \in \Exp\), if \(e_1 \bisim e_2\), then \(\T^* \vdash e_1 = e_2\).
\end{restatable}

\section{Examples and Nonexamples}%
\label{sec:examples}

As we have hinted at, the theory of semilattices \(\SL\), guarded algebra \(\GA\), and convex algebra \(\CA\) are supported and malleable, and therefore fit our framework.
But, as we have already seen, some equational theories do not admit support, and others are not malleable.
In this section, we discuss the scope of our story.

The following result gives a sufficient condition for malleability.

\begin{restatable}{proposition}{propMalleableSufficient}%
    \label{prop:sufficient malleable}
    Let \(\T\) be an equational theory for a signature \(S\) consisting of constants and binary operations.
    \(\T\) is malleable if both of the following hold:
    \begin{enumerate}
        \item \emph{Skew commutativity.} For any binary operation \(\sigma \in S\), there is a binary operation \(\tau \in S\) such that \(\T \vdash \sigma(x, y) = \tau(y, x)\).
        \item \emph{Skew associativity.} For any binary operations \(\sigma_1,\sigma_2 \in S\), there are binary operations \(\tau_1,\tau_2 \in S\) such that \(\T \vdash \sigma_1(x, \sigma_2(y, z)) = \tau_1(\tau_2(x, y), z)\).
    \end{enumerate}
\end{restatable}

It is immediate from the axioms of \(\SL\), \(\GA\), and \(\CA\) that all three satisfy these properties.
However, some malleable theories are not skew-associative.

\begin{instantiation}%
    \label{inst:probgkat}
    The theory of \emph{guarded convex algebra} \(\GC\) consists of one constant symbol \(0\), one binary operation \(+_b\) for each Boolean expression \(b \in \mathit{BA}\) (see Inst.~\ref{inst:guarded alg}), one binary operation \(\oplus_p\) for each \(p \in [0,1]\), and is axiomatized by the equations of \(\GA\), \(\CA\), and the \emph{distribution law} \(x \oplus_p (y +_b z) = (x \oplus_p y ) +_b (x \oplus_p z)\).
    Its free-algebra construction is given by \((\D(\bot + (-))^\At, \eta, \rho)\) where \(\eta_X(x)(\alpha) = \delta_x\); \((\chi_1 +_b^\rho \chi_2)(\alpha)(x) = \chi_1(\alpha)(x)\) if $\alpha \leq b$, and \((\chi_1 +_b^\rho \chi_2)(\alpha)(x) = \chi_2(\alpha)(x)\) otherwise; \((\chi_1 \oplus_p^\rho \chi_2)(\alpha)(x) = p\chi_1(\alpha)(x) + (1-p)\chi_2(\alpha)(x)\), and $0^\rho(\alpha) = \delta_\bot$.

    Guarded convex algebra is a supported malleable theory that is not skew associative.
    Indeed, we can take \(\supp \colon \D(\bot + (-))^\At \Rightarrow \P\) to be \(\supp_X(\chi) = \bigcup_{\alpha \in \At}\{x \in X \mid \chi(\alpha)(x) > 0\}\).
    It is not difficult to show that this is a natural transformation that satisfies the requirements of a support.
    To see why \(\GC\) is not skew-associative, consider the term \(x +_{0.5} (y +_b z)\).
    A simple case analysis reveals that it is not equivalent to \(\tau_1(\tau_2(x, y), z)\) for any binary operations $\tau_1$ and $\tau_2$.

    We illustrate the proof that \(\GC\) is malleable in the special case of \(X = \{x, y, z\}\) and \(\At = \{\alpha_1, \alpha_2\}\).
    Consider \(\chi = \theta_1 +_{\alpha_1} \theta_2\) for some probability distributions \(\theta_1,\theta_2\) on \(X\).
    The convex algebra \(\D(\bot + X)\) can be visualized as the \(3\)-simplex in \(\mathbb R^4\)~\cite{sokolova-woracek-15} with extremal points \(\delta_x, \delta_y, \delta_z, \delta_\bot\).
    We only need one of its faces, the convex hull of \(\delta_x, \delta_y, \delta_z\), depicted in~\eqref{img:gc malleable}.
    There, \(\chi\) represents two points, one for each of \(\alpha_1,\alpha_2\).
    To obtain terms \(s(u, v)\), \(t_1(x, y)\) and \(t_2(z)\) such that \(\chi = s^\rho(t_1^\rho, t_2^\rho)\), draw straight lines from \(\delta_z\) through \(\theta_1\) to the segment between \(\delta_x\) and \(\delta_y\).

    \noindent
    \begin{minipage}{0.65\textwidth}
        \ \ The endpoints of the drawn lines represent distributions obtained from terms of the form \(r_1(x,y),r_2(x,y)\), i.e., \(\theta_1' = r_1^\rho(x, y)\) and \(\theta_2' = r_2^\rho(x, y)\).
        Then \(\chi = (\theta_1' \oplus_p \delta_z) +_{\alpha_1} (\theta_2' \oplus_q \delta_z)\) for some \(p,q \in [0,1]\).
        If we choose \(s(u, v) = (u \oplus_p v) +_{\alpha_1} (u \oplus_q v)\), \(t_1(x, y) = r_1(x, y) +_{\alpha_1} r_2(x, y)\) and \(t_2 = z\), then \(\chi = s^\rho(t_1^\rho, t_2^\rho)\).
    \end{minipage}
    \begin{minipage}{0.32\textwidth}
        \vspace{-1em}
        \begin{equation}
            \label{img:gc malleable}
            \begin{tikzpicture}
                \node (0) at (0,0) {
                    \includegraphics[scale=0.35]{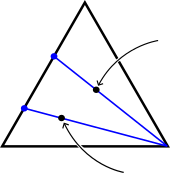}
                };
                \node (2) at (0,1.1) {\(\delta_x\)};
                \node (3) at (-1.1,-0.6) {\(\delta_y\)};
                \node (1) at (1.1,-0.6) {\(\delta_z\)};

                \node (5) at (0.9, 0.4) {\(\theta_1\)};
                \node (5) at (0.6, -0.9) {\(\theta_2\)};
                \node (5) at (-0.55, 0.5) {\(\theta_1'\)};
                \node (5) at (-0.9, -0.1) {\(\theta_2'\)};
            \end{tikzpicture}
        \end{equation}
    \end{minipage}
\end{instantiation}

\emph{Guarded convex algebra} is the equational theory \(\GC\) underlying the recently introduced \emph{probabilistic guarded Kleene algebra with tests} (or \emph{ProbGKAT})~\cite{rozowski-etal-2023,schmid-thesis}.
However, the skip-free universal-star fragment of \(\GC\) is not the obvious skip-free fragment of ProbGKAT, because the latter allows only the binary stars \(e_1^{(s)}e_2\) for \(s(u, v) = u +_b v\) and \(s = u \oplus_p v\).
In contrast, skip-free unified-star expressions for \(\GC\) allow \emph{mixed loops}, like $e_1^{(s)}e_2$ with $s = u \oplus_p (u +_b v)$, which enters the loop body with probability $p$, and with probability $1-p$ does this iff $b$ is true.


\newcommand{\Semi}{\mathbb{S}}
\newcommand{\Mod}{{\mathsf{Mod}}}


\begin{instantiation}
    A \emph{(unital) semiring} is a set \(\Semi\) equipped with two constants \(0\) and \(1\) and two binary operations \(+\) and \(\times\) (written as juxtaposition) such that \((\Semi, +, 0)\) is a commutative monoid, \((\Semi, \times, 1)\) is a monoid, and the distributive laws \(p (q + r) = p q + p r\) and \((p + q) r = pr + qr\) hold.
    The theory \(\Semi\Mod\) of \emph{semimodules} over a semiring \(\Semi\) has a signature consisting of a constant \(0\), a binary operation \(\oplus\), and a unary operation \(p \cdot (-)\) for each \(p \in \Semi\).
    The axioms of \(\Semi\Mod\) state that \(\oplus\) is commutative, associative, and has \(0\) as a neutral element (the commutative monoid axioms), as well as
    \(0 \cdot x = 0\),
    \(1 \cdot x = x\),
    \(p \cdot (q \cdot x) = (pq) \cdot x\),
    \(p\cdot(x \oplus y) = (p\cdot x) \oplus (p \cdot y)\), and
    \((p + q) \cdot x = (p \cdot x) \oplus (q \cdot x)\),
    for any \(p,q \in \Semi\).

    Given a function \(\theta \colon X \to \Semi\), define \(\supp_X(\theta) = \{x \in X \mid \theta(x) \neq 0\}\).
    The free algebra construction for \(\Semi\Mod\) is given by \((\mathcal O_\Semi, \eta, \rho)\), where \(\mathcal O_\Semi X = \{\theta \colon X \to \Semi \mid \supp_X(\theta)\text{ is finite}\}\); \(\eta(x)(y) = \textbf{if \(x = y\) then \(1\) else \(0\)}\); and where
    \(
        0^\rho(x) = 0
    \),
    \(
        (\theta_1 \oplus^\rho \theta_2)(x) = \theta_1(x) + \theta_2(x)
    \), and
    \(
        (p \cdot^\rho \theta)(x) = p\theta(x)
    \).
    An \(\mathcal O_\Semi\)-system is essentially a weighted transition system with weights that live in \(\Semi\).

    The theory \(\Semi\Mod\) is supported malleable: 
    \(\supp_X(\theta)\) is finite for each \(\theta \in \mathcal O_\Semi X\) by definition, so we obtain a natural transformation \(\supp \colon \mathcal O_\Semi \Rightarrow \P\) that clearly satisfies the requirements of a support for \(\Semi\Mod\).
    To see malleability, observe that up to \(\Semi\Mod\), every term \(t(\vec x, \vec y)\) with disjoint \(\vec x,\vec y\) is equivalent to one of the form \([(p_1 \cdot x_1) \oplus \cdots \oplus (p_n \cdot x_n)] \oplus [(q_1 \cdot y_1) \oplus \cdots \oplus (q_m \cdot y_m)]\) for some \(p_i,q_j\in \Semi\).

    %
\end{instantiation}

\paragraph*{An Unfortunate Nonexample.}
Several authors have taken an interest in mixing nondeterminism with probability~\cite{bonchi-sokolova-vignudelli-21,keimel-plotkin-16,liellcock-staton-24,mislove-aknine-04,varacca-winskel-06}.
A natural choice for the underlying equational theory in this case is the theory of \emph{convex semilattices} \(\CS\)~\cite{bonchi-sokolova-vignudelli-21}, which consists of \(\SL\), \(\CA\) and the distributive law \(x \oplus_p (y + z) = (x \oplus_p y) + (x \oplus_p z)\).
The free-algebra construction for \(\CS\) is \((\mathcal C, \eta, \rho)\) where \(\mathcal CX\) is the set of convex subsets of \(\D(\bot + (-))\) that include \(\delta_\bot\), \(\eta_X(x) = \{p\delta_x + (1-p)\delta_\bot \mid p \in [0,1]\}\), \(0^{\rho_X} = \{\delta_\bot\}\), \(U \oplus_p^{\rho_X} V = \{p\theta_1 + (1-p)\theta_2 \mid \theta_1 \in U,\theta_2 \in V\}\), and \(U +^{\rho_X} V = \conv(U \cup V)\) is the convex hull of \(U \cup V\)~\cite{bonchi-sokolova-vignudelli-21} (see also~\cite[Example 4.1.14]{schmid-thesis}).

The theory of convex semilattices admits the obvious support but, despite the similarity to \(\GC\), it is not malleable.
Indeed, there are no terms \(s(u, v)\), \(t_1(x, y)\), and \(t_2(z)\) such that \(\CS \vdash s(t_1(x, y), t_2(z)) = x + (y \oplus_{\frac12} z)\).
To see why, recall that the space of probability distributions on \(X = \{x,y,z\}\) can be identified with a face of the \(3\)-simplex, depicted as a black triangle in~\eqref{img:cs not malleable 1}.
\begin{equation}
    \label{img:cs not malleable 1}
    \begin{gathered}
        \begin{tikzpicture}[scale=0.75]
            \node (0) at (0,0) {
            \includegraphics[scale=0.3]{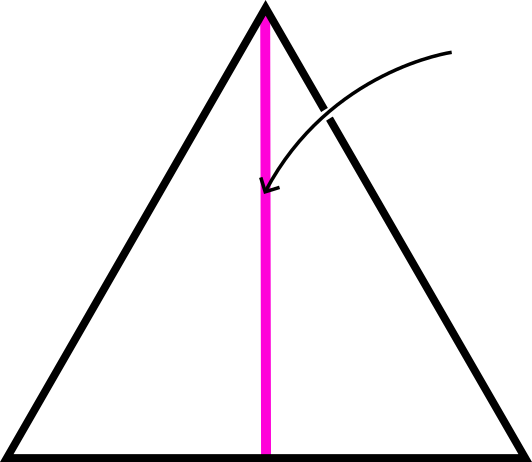}
            };
            \node (2) at (0,1) {\(\delta_x\)};
            \node (3) at (-1.1,-0.7) {\(\delta_y\)};
            \node (1) at (1.1,-0.7) {\(\delta_z\)};
            \node (form) at (2.7,0.5) {\(\delta_x +^{\rho_X} (\delta_y \oplus_{\frac12}^{\rho_X} \delta_z)\)};
        \end{tikzpicture}
    \end{gathered}
    \qquad
    \begin{gathered}
        \begin{tikzpicture}[scale=0.75]
            \node (0) at (0,0) {
                \includegraphics[scale=0.3]{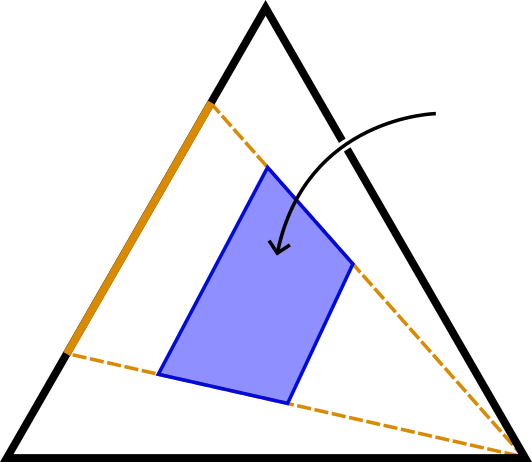}
            };
            \node (2) at (0,1) {\(\delta_x\)};
            \node (3) at (-1.1,-0.7) {\(\delta_y\)};
            \node (1) at (1.1,-0.7) {\(\delta_z\)};
            \node (form) at (2.9,0.5) {\(s^{\rho_X}(t_1^{\rho_X}(x, y), t_2^{\rho_X}(z))\)};
        \end{tikzpicture}
    \end{gathered}
\end{equation}
The line down the middle of the left triangle in~\eqref{img:cs not malleable 1} is the convex set of probability distributions that corresponds to the term \(x + (y \oplus_{\frac12} z)\) (cf.~\cite[Fig. 1]{pirog-staton-17}).
The highlighted region in the center of the right triangle is the general shape that any convex set of the form \(s(t_1(x, y), t_2(z))\) must take.
No convex set of this form is equal to the line segment in~\eqref{img:cs not malleable 1}.
Thus, \(\CS\) is not malleable.

On the other hand, the theory of convex semilattices still obtains a syntax of skip-free unified-star expressions, a bisimilarity semantics, and a sound set of axioms \(\CS^*\) from the framework presented in the paper.
So, we ask: is \(\CS^*\) complete with respect to bisimilarity of skip-free unified-star expressions?

\section{Discussion and Future Work}

Given a supported and malleable equational theory $\T$, we can derive a notion of bisimilarity and a complete axiomatization for ``skip-free'' processes with $\T$-branching.
This framework recovers existing completeness theorems for $\SL^*$ (the result by Grabmayer and Fokkink~\cite{grabmayer-fokkink-2020}) and $\GA^*$ (skip-free GKAT up to bisimilarity~\cite{kappe-schmid-silva-2023}), and yields new ones for $\CA^*$ ($1$-free probabilistic regular expressions up to bisimilarity~\cite{rozowskiS24}) and $\GC^*$ (a slightly generalized skip-free ProbGKAT~\cite{rozowski-etal-2023}).

We would like our framework to abstract the completeness theorem of regular expressions up to bisimilarity~\cite{grabmayer-2022}.
This could settle the open completeness problem for full GKAT up to bisimilarity~\cite{schmid-kappe-silva-2021}, and possibly its trace semantics~\cite{smolkaFHKKS-20}.

The theories we consider all contain a constant $0$, which stands for the \emph{deadlocked} process, which does not allow any branching and satisfies $0e = 0$.
It would be interesting to see what would be necessary to guarantee that $e0 = 0$.
This would make $0$ act like the ``predictable failure'' studied by Baeten and Bergstra in~\cite{baeten-bergstra-90}.
Earlier work in the setting of GKAT has shown that completeness for this extended system can be derived from the original~\cite{schmid-kappe-silva-2021,kappe-schmid-silva-2023}.

Unified star-expressions give a middle ground between a Kleene-star and general recursion, by focusing on loops that derive from $\T$.
We would also like to investigate the hierarchy of expressiveness of star-expressions for $n$-ary operators.
Perhaps such an extension would help to get a completeness theorem for $\CS^*$.

A natural question to ask is which equational theories are supported malleable.
For example, all of our examples are skew commutative (in the sense of \cref{prop:sufficient malleable}), but it is currently not clear if this is a necessary condition.
Also, conspicuously, the distributive law in \(\GC\) allowed us to mix \(\GA\) with \(\CA\) to produce a malleable theory, while the distributive law in \(\CS\) did not.
We would like know if this is related to the existence of a distributive law of monads \(\D\R \Rightarrow \R\D\) and the lack of a distributive law \(\D\P \Rightarrow \P\D\)~\cite{varacca-winskel-06}, or at least to the \emph{composite theories} of~\cite{pirog-staton-17} that guarantee the existence of a distributive law.

Finally, because bisimilarity coincides with provable equivalence for the equational theories satisfying our constraints, it is also a congruence.
We wonder whether this implies the existence of a distributive law
\`a la Turi and Plotkin~\cite{turi-plotkin-97}.

\paragraph*{Acknowledgements}
T.~Kappé was partially supported by the European Union’s Horizon 2020 research and innovation programme under grant no.\ 101027412 (VERLAN), and partially by the Dutch research council (NWO) under grant no.\ VI.Veni.232.286 (ChEOpS).

\bibliographystyle{splncs04}
\bibliography{main.bib}

\ifarxiv%
\appendix

\input{appendix.tex}
\fi

\end{document}

%% file: appendix.tex

\clearpage
\appendix

\section{Proof of \cref{prop:Exp locally finite} and Soundness (\cref{thm:soundness})}

\propLocallyFinite*

\begin{proof}
    The proof roughly follows~\cite[Lemma 4.4.5]{schmid-thesis}.
    We start by recalling a result of Gumm and Schr\"oder~\cite[Theorem 3.1]{gumm-schroeder-02}: for coalgebras of endofunctors on \(\Set\), the intersection of two subcoalgebras is always a subcoalgebra.
    Because $M$-systems are coalgebras, it suffices to construct a finite subsystem $(U(e), \gamma_{U(e)})$ of $(\Exp, \gamma)$ containing $e$, as we can intersect all (finitely many) of subsystems of $(U(e), \gamma_{U(e)})$ that contain $e$ to find the smallest subsystem containing $e$.

    Let \(e \in \Exp\).
    We start by defining \(U \colon \Exp \to \P(\Exp)\) inductively, as follows:
    \begin{gather*}
        U(a) = \{a\}
        \qquad
        U(\sigma(e_1, \dots, e_n)) = \{\sigma(e_1, \dots, e_n)\} \cup \bigcup U(e_i)
        \\
        U(e_1e_2) = \{fe_2 \mid f \in U(e_1)\} \cup U(e_2)
        \\
        U(e_1^{(s)}e_2) = \{e_1^{(s)}e_2\} \cup \bigcup \{f(e_1^{(s)}e_2) \mid f \in U(e_1)\} \cup U(e_2)
    \end{gather*}
    By induction, \(U(e)\) is finite and \(e \in U(e)\) for all \(e \in \Exp\).
    It remains to show that $U(e)$ is a subsystem, i.e., that we can restrict $\gamma$ to $\gamma_{U(e)}\colon U(e) \to B_M U(e)$, which we do by means of the following two claims.

    \begin{claim}
    For any \(e \in \Exp\), there is a term \(t((\vec b, \checkmark), (\vec a, \vec f)) \in S^*(\Act \times (\checkmark + U(e)))\) such that \(\gamma(e) = t^\rho\).
    In particular, this means that \(f_i \in U(e)\) for each \(i\).
    \end{claim}
    \begin{proof}[of claim]
    \begin{itemize}
        \item Let \(e = a \in \Act\).
        Here, \(\gamma(a) = (a, \checkmark)^\rho\), so the claim is vacuous.

        \item Let \(e = \sigma(e_1, \dots, e_n)\).
        Using the IH, let \(t_i \in S^*(\Act \times (\checkmark + U(e_i)))\) such that \(\gamma(e_i) = t_i^\rho\).
        Then \(\gamma(e) = \sigma(\gamma(e_1), \dots, \gamma(e_n)) = \sigma(t_1, \dots, t_n)^\rho\), and \(\sigma(t_1, \dots, t_n) \in S^*(\Act \times (\checkmark + U(e)))\) since \(\bigcup_{i=1}^n U(e_i) \subseteq U(e)\).

        \item Let \(e = e_1e_2\).
        Using the IH, let \(t_1 = t_1((\vec b, \checkmark), (\vec a, \vec f)) \in S^*(\Act \times (\checkmark + U(e_1)))\) be such that \(\gamma(e_1) = t_1^\rho\).
        Then \[
            \gamma(e_1e_2) = t_1((\vec b, e_2), (\vec a, \vec fe_2)) \in S^*(\Act \times (\checkmark + U(e)))
        \]
        because \(f_i e_2 \in U(e_1e_2)\) for each \(i\).

        \item Let \(e = e_1^{(s)} e_2\).
        By induction, we find \(t_k = t_k((\vec b, \checkmark), (\vec a, \vec f)) \in S^*(\Act \times (\checkmark + U(e_k)))\) such that \(\gamma(e_k) = t_k^\rho\) for $k \in \{1,2\}$.
        Then
        \[
            s\Big(t_1((\vec b, e_1^{(s)}e_2), (\vec a, \vec f(e_1^{(s)}e_2))), t_2((\vec d, \checkmark), (\vec c, \vec g))\Big)  \in S^*(\Act \times (\checkmark + U(e)))
        \]
        because \(e_1^{(s)}e_2, f_i(e_1^{(s)}e_2), g_j \in U(e_1^{(s)}e_2)\) for each \(i, j\), and the interpretation of the term above is equal to $\gamma(e_1^{(s)}e_2)$ by definition.
    \end{itemize}
    \end{proof}

    \begin{claim}
    Let \(e,f,g \in \Exp\). If \(g \in U(f)\) and \(f \in U(e)\), then \(g \in U(e)\).
    \end{claim}
    \begin{proof}[of claim]
    By induction on \(e\).
    \begin{itemize}
        \item If \(e = a \in \Act\), the claim amounts to \(g \in U(f)\) and \(f \in U(e) = \{a\}\), so \(f = a\) and \(g \in U(f) = U(a) = U(e)\).

        \item If \(e = \sigma(\vec e)\), then \(f \in U(e_i)\) for some \(i\).
        By the induction hypothesis, \(g \in U(e_i) \subseteq U(e)\).

        \item If \(e = e_1e_2\), then either (i) \(f \in U(e_2)\), or (ii) \(f = e_1'e_2\) for some \(e_1' \in U(e_1)\).
        In case (i), the induction hypothesis then implies that \(g \in U(e_2) \subseteq U(e_1e_2)\).
        In case (ii), \(g \in U(e_1'e_2)\), so either \(g \in U(e_2) \subseteq U(e_1e_2)\) or \(g = e_1''e_2\) for some \(e_2'' \in U(e_1')\).
        In the latter situation, by the induction hypothesis, \(e_1'' \in U(e_1)\), so \(g = e_1''e_2 \in U(e_1e_2)\).

        \item If \(e = e_1^{(s)}e_2\), then either (i) \(f \in U(e_2)\), or (ii) \(f = e_1'(e_1^{(s)}e_2)\) for some \(e_1' \in U(e_1)\).
        In case (i), the induction hypothesis then implies that \(g \in U(e_2) \subseteq U(e_1e_2)\).
        In case (ii), \(g \in U(e_1'(e_1^{(s)}e_2))\), so either \(g \in U(e_1^{(s)}e_2)\) or \(g = e_1''(e_1^{(s)}e_2)\) for some \(e_2'' \in U(e_1')\).
        In the latter situation, by the induction hypothesis, \(e_1'' \in U(e_1)\), so \(g = e_1''(e_1^{(s)}e_2) \in U(e_1^{(s)}e_2)\).
    \end{itemize}
    This establishes the claim.
    \end{proof}

    Together, these two claims imply that if $f \in U(e)$, then $\gamma(e) \in B_M U(e)$.
    To see this, suppose $f \in U(e)$; by the first claim, there exists a $t \in S^*(\Act \times (\checkmark + U(f)))$ such that $\gamma(f) = t^\rho$.
    By the second claim, $U(f) \subseteq U(e)$, and so $t \in S^*(\Act \times (\checkmark + U(e)))$, which means that $\gamma(f) \in B_M U(e)$.
    Hence, $\gamma$ can be restricted to $U(e)$, meaning $U(e)$ is a subsystem of $(\Exp, \gamma)$ containing $e$.
\end{proof}

The next theorem we prove is the soundness theorem.
We need the following lemma, which also appears as a part of \cref{prop:solutions are homoms}.

\begin{lemma}%
    \label{lem:soundness lemma}
    There is a unique \(M\)-system structure \((\Exp/{\equiv}, [\gamma]_\equiv)\) on the set of provable equivalence classes \(\Exp/{\equiv}\) such that \([-]_\equiv \colon (\Exp, \gamma) \to (\Exp/{\equiv}, [\gamma]_\equiv)\).
\end{lemma}

\begin{proof}
    We aim to find a map \([\gamma]_\equiv \colon \Exp/{\equiv} \to B_M(\Exp/{\equiv})\) that makes the square below commute.
    \begin{equation}
        \label{eq:diagonal fill-in}
        \begin{tikzcd}
            \Exp \ar[d, "\gamma"] \ar[r, "{[-]_\equiv}"]
            & {\Exp/{\equiv}} \ar[d, dashed, "{[\gamma]_\equiv}"] \\
            B_M\Exp \ar[r, "{B_M([-]_\equiv)}"]
            & B_M{\Exp/{\equiv}}
        \end{tikzcd}
    \end{equation}
    Since \([-]_\equiv\) is surjective, there can be at most one map that makes~\eqref{eq:diagonal fill-in} commute. 
    Thus, to find such a map is to find the unique one.
    We proceed with the construction of \([\gamma]_\equiv\) using a \emph{diagonal fill-in} argument.

    \begin{claim}
    If \(e,f\) are identified by \([-]_\equiv\), i.e., \(\T^* \vdash e = f\), then \(e,f\) are identified by \(B_M([-]_\equiv) \circ \gamma\).
    \end{claim}

    \begin{proof}[of claim]
    We proceed by induction on the derivation of \(\T^*\vdash e = f\).
    \begin{itemize}
        \item (\(\T\)) Suppose \(\T \vdash t_1(\vec x) = t_2(\vec x)\) and \(e = t_1(\vec e)\) and \(f = t_2(\vec e)\).
        In this case, \(\gamma(e) = \gamma(f)\) because $M(\Act \times (\checkmark + \Exp))$ has the structure of a $\T$-algebra, so \(B_M([-]_\equiv) \circ \gamma(e) = B_M([-]_\equiv) \circ \gamma(f)\) follows.

        \item (A) Now suppose \(e = e_1(e_2e_3)\) and \(f = (e_1e_2)e_3\).
        Let \(\gamma(e_1) = t((\vec b, \checkmark), (\vec a, \vec f))\).
        \begin{align*}
            B_M([-]_\equiv) \circ \gamma(e)
            &= B_M([-]_\equiv)(t^\rho((\vec b, e_2e_3), (\vec a, \vec f(e_2e_3)))) \\
            &= t^\rho((\vec b, e_2e_3), (a_1, [f_1(e_2e_3)]_\equiv), \dots, (a_n, [f_n(e_2e_3)]_\equiv)) \\
            &= t^\rho((\vec b, e_2e_3), (a_1, [(f_1e_2)e_3]_\equiv), \dots, (a_n, [(f_n e_2)e_3]_\equiv)) \\
            &= \hspace{-2pt} B_M([-]_\equiv)(t^\rho((\vec b, e_2e_3), (a_1, (f_1e_2)e_3), \dots, (a_n, (f_n e_2)e_3))) \\
            &= B_M([-]_\equiv) \circ \gamma((e_1e_2)e_3)
        \end{align*}

        \item (D) Now consider the case of \(e = t(\vec e)g\) and \(f = t(\vec eg)\).
        Let \[
            \gamma(e_i) = t_i((\vec b_{i}, \checkmark), (\vec a_{i}, \vec f_{i}))
        \]
        for each \(i\) and observe that we have both
        \[
            \gamma(e_i g) = t_i((\vec b_{i}, g), (\vec a_{i}, \vec f_{i}g))
        \]
        for each \(i\) as well as
        \[
            \gamma(t(\vec e))
            = t^\rho\Big(
                t_1^\rho((\vec b_{1}, \checkmark), (\vec a_{1}, \vec f_{1})),
                \dots,
                t_n^\rho((\vec b_{n}, \checkmark), (\vec a_{n}, \vec f_{n}))
            \Big)
        \]
        Then we can proceed with the following calculation.
        \begin{align*}
            \gamma(e)
            &= \gamma(t(\vec e)g)  \\
            &=
                t^\rho(
                    t_1^\rho((\vec b_{1}, g), (\vec a_{1}, \vec f_{1}g)),
                    \dots,
                    t_n^\rho((\vec b_{n}, g), (\vec a_{n}, \vec f_{n}g))
                )
            \\
            &= \gamma(t(\vec e g)) \\
            &= \gamma(f)
        \end{align*}
        Again, by extension, \(B_M([-]_\equiv) \circ \gamma(e) = B_M([-]_\equiv) \circ \gamma(f)\).

        \item (U) Now consider \(e = e_1^{(s)}e_2\) and \(f = s(e_1(e_1^{(s)}e_2), e_2)\).
        If we let \(\gamma(e_1) = t((\vec b, \checkmark), (\vec a, \vec f))\), then we can derive
        \begin{align*}
            \gamma(e_1^{(s)}e_2)
            &= s^\rho\Big(t^\rho((\vec b, e_1^{(s)}e_2), (\vec a, \vec f(e_1^{(s)}e_2))), \gamma(e_2)\Big) \\
            &= s^\rho\Big(\gamma(e_1(e_2^{(s)}e_2)), \gamma(e_2)\Big) \\
            &= \gamma(s(e_1(e_1^{(s)}e_2), e_2))
        \end{align*}
        Again, by extension, \(B_M([-]_\equiv) \circ \gamma(e) = B_M([-]_\equiv) \circ \gamma(f)\).

        \item (RSP) Suppose that \(e = g\), \(f = e_1^{(s)}e_2\), that \(\T^* \vdash g = s(e_1g, e_2)\) --- so in particular $\T^* \vdash g = e_1^{(s)}e_2$ by (RSP) --- and for an induction hypothesis assume \(B_M([-]_\equiv)\circ \gamma(g) = B_M([-]_\equiv)\circ \gamma(s(e_1g, e_2))\).
        Let \(\gamma(e_1) = t^\rho((\vec b, \checkmark), (\vec a, \vec f))\).
        We carry out the following calculation.
        \begin{align*}
            B_M([-]_\equiv) \circ \gamma(g)
            &= B_M([-]_\equiv)\circ \gamma(s(e_1g, e_2)) \\
            &= B_M([-]_\equiv)(s^\rho(\gamma(e_1g), \gamma(e_2)) ) \\
            &= B_M([-]_\equiv)(s^\rho(t^\rho((\vec b, g), (\vec a, \vec fg)), \gamma(e_2)) ) \\
            &= s^\rho(t^\rho((\vec b, [g]_\equiv), (\vec a, [\vec fg]_\equiv)), \gamma(e_2)) \\
            &= s^\rho(t^\rho((\vec b, [e_1^{(s)}e_2]_\equiv), (\vec a, [\vec f(e_1^{(s)}e_2)]_\equiv)), \gamma(e_2)) \tag{RSP} \\
            &= s^\rho(B_M([-]_\equiv) \circ \gamma(e_1(e_1^{(s)}e_2)), \gamma(e_2)) \\
            &= B_M([-]_\equiv) \circ \gamma(s(e_1(e_1^{(s)}e_2), e_2)) \\
            &= B_M([-]_\equiv) \circ \gamma(e_1^{(s)}e_2) \tag{previous case}
        \end{align*}
        \item
        (Congruence, $\sigma$)
        Suppose that $e = \sigma(e_1, \dots, e_n)$ and $f = \sigma(f_1, \dots, f_n)$ such that for all $i$ we have $\T^* \vdash e_i = f_i$.
        By induction, we then know that for all $i$ we have $B_M([-]_\equiv) \circ \gamma(e_i) = B_M([-]_\equiv) \circ \gamma(f_i)$.
        We then derive:
        \begin{align*}
        B_M([-]_\equiv) \circ \gamma(e)
            &= B_M([-]_\equiv) \circ \gamma(\sigma(e_1, \dots, e_n)) \\
            &= B_M([-]_\equiv)(\sigma^\rho(\gamma(e_1), \dots, \gamma(e_n))) \\
            &= \sigma^\rho(B_M([-]_\equiv)(\gamma(e_1)), \dots, B_M([-]_\equiv)(\gamma(e_n))) \\
            &= \sigma^\rho(B_M([-]_\equiv)(\gamma(f_1)), \dots, B_M([-]_\equiv)(\gamma(f_n))) \\
            &= B_M([-]_\equiv)(\sigma^\rho(\gamma(f_1), \dots, \gamma(f_n))) \\
            &= B_M([-]_\equiv) \circ \gamma(\sigma(f_1, \dots, f_n)) \\
            &= B_M([-]_\equiv) \circ \gamma(f)
        \end{align*}
        \item
        (Congruence, concatenation)
        Suppose that $e = e_1e_2$ and $f = f_1f_2$ such that for both $i$ we have that $\T^* \vdash e_i = f_i$.
        By induction, we then know that for both $i$ we have $B_M([-]_\equiv) \circ \gamma(e_i) = B_M([-]_\equiv) \circ \gamma(f_i)$.
        Now, if $\gamma(e_1) = t^\rho((\vec a, \checkmark), (\vec b, \vec g))$ and $\gamma(f_1) = s^\rho((\vec c, \checkmark), (\vec d, \vec h))$, then $t^\rho((\vec a, \checkmark), (\vec b, [\vec g]_\equiv)) = s^\rho((\vec c, \checkmark), (\vec d, [\vec h]_\equiv))$, so via the substitution
        \begin{mathpar}
        (a_i, \checkmark) \mapsto (a_i, [e_2]_\equiv)
        \and
        (b_i, [g_i]_\equiv) \mapsto (b_i, [g_i e_2]_\equiv)
        \\
        (c_i, \checkmark) \mapsto (c_i, [f_2]_\equiv)
        \and
        (d_i, [h_i]_\equiv) \mapsto (c_i, [h_i f_2]_\equiv)
        \end{mathpar}
        (which is well-defined because $e_2 \equiv f_2$) we find that
        \[
            t^\rho((\vec a, [e_2]_\equiv), (\vec b, [\vec ge_2]_\equiv)) = s^\rho((\vec c, [f_2]_\equiv), (\vec d, [\vec h f_2]_\equiv))
        \]
        With this in hand, we can derive
        \begin{align*}
        B_M([-]_\equiv) \circ \gamma(e_1e_2)
            &= B_M([-]_\equiv)(t^\rho((\vec{a}, e_2), (\vec{b}, \vec{g}e_2))) \\
            &= t^\rho((\vec{a}, [e_2]_\equiv), (\vec{b}, [\vec{g}e_2]_\equiv)) \\
            &= s^\rho((\vec{c}, [f_2]_\equiv), (\vec{d}, [\vec{h}f_2]_\equiv)) \\
            &= B_M([-]_\equiv)(s^\rho((\vec{c}, f_2), (\vec{b}, \vec{h}f_2))) \\
        \end{align*}
        \item
        (Congruence, star rule)
        Suppose that $e = e_1^{(s)}e_2$ and $f = f_1^{(s)}f_2$ for some term $s = s(u, v)$, and that for both $i$, we have that $\T^* \vdash e_i = f_i$.
        By induction, we then know that $B_M([-]_\equiv) \circ \gamma(e_i) = B_M([-]_\equiv) \circ \gamma(f_i)$.
        Now, if $\gamma(e_1) = t_e^\rho((\vec a, \checkmark), (\vec b, \vec g))$ and $\gamma(f_1) = t_f^\rho((\vec c, \checkmark), (\vec d, \vec h))$, then $t_e^\rho((\vec a, \checkmark), (\vec b, [\vec g]_\equiv)) = t_f^\rho((\vec c, \checkmark), (\vec d, [\vec h]_\equiv))$, so via the substitution
        \begin{mathpar}
        (a_i, \checkmark) \mapsto (a_i, [e_1^{(s)}e_2]_\equiv)
        \and
        (b_i, [g_i]_\equiv) \mapsto (b_i, [g_i (e_1^{(s)}e_2)]_\equiv)
        \\
        (c_i, \checkmark) \mapsto (c_i, [f_1^{(s)}f_2]_\equiv)
        \and
        (d_i, [h_i]_\equiv) \mapsto (c_i, [h_i f_1^{(s)}f_2]_\equiv)
        \end{mathpar}
        (which is well-defined because $e_1^{(s)}e_2 \equiv f_1^{(s)}f_2$), we find that
        \[
            t_e^\rho((\vec a, [e_1^{(s)}e_2]), (\vec b, [\vec g (e_1^{(s)}e_2)]_\equiv))
                = t_f^\rho((\vec c, [f_1^{(s)}f_2]), (\vec d, [\vec h (f_1^{(s)}f_2)]_\equiv))
        \]
        With this in mind, we can derive as follows:
        \begin{align*}
        B_M([-]_\equiv) \circ \gamma(e_1^{(s)}e_2)
            &= B_M([-]_\equiv)(s^\rho(t_e^\rho((\vec{a}, e_1^{(s)}e_2), (\vec{b}, \vec{g}(e_1^{(s)}e_2))), \gamma(e_2))) \\
            &= s^\rho(t_e^\rho((\vec{a}, [e_1^{(s)}e_2]_\equiv), (\vec{b}, [\vec{g}(e_1^{(s)}e_2)]_\equiv)), \\
            &\hspace{1cm} B_M([-]_\equiv) \circ \gamma(e_2)) \\
            &= s^\rho(t_f^\rho((\vec{a}, [f_1^{(s)}f_2]_\equiv), (\vec{b}, [\vec{h}(f_1^{(s)}f_2)]_\equiv)), \\
            &\hspace{1cm} B_M([-]_\equiv) \circ \gamma(f_2)) \\
            &= B_M([-]_\equiv)(s^\rho(t_f^\rho((\vec{a}, f_1^{(s)}f_2), (\vec{b}, \vec{g}(f_1^{(s)}f_2))), \gamma(f_2))) \\
            &= B_M([-]_\equiv) \circ \gamma(f_1^{(s)}f_2)
        \end{align*}
    \end{itemize}
    This concludes the proof of the claim.
    \end{proof}

    The \(M\)-system structure can now be given by the formula
    \[
        [\gamma]_\equiv ([e]_\equiv) = B_M([-]_\equiv) \circ \gamma(e)
    \]
    Note that this automatically ensures that~\eqref{eq:diagonal fill-in} commutes, i.e., \([-]_\equiv\) is a homomorphism of \(M\)-systems.
\end{proof}

\thmSoundness*

\begin{proof}
    We appeal to \cref{lem:soundness lemma}.
    If \(\T^* \vdash e = f\), then \([e]_\equiv = [f]_\equiv\).
    Since \([-]_\equiv\) is a homomorphism of \(M\)-systems, \(e \bisim f\).
\end{proof}

\section{Proofs of \cref{prop:solutions are homoms} and \cref{prop:solution pullbacks}}

The first part of \Cref{prop:solutions are homoms} is \cref{lem:soundness lemma}.
For the second part, we need to establish the following result.

\begin{lemma}%
    \label{lem:fundamental}
    Let \(e \in \Exp\) and \(\gamma(e) = t^\rho((\vec b, \checkmark), (\vec a, \vec e))\).
    Then
    \begin{equation}
        \T^* \vdash e = t(\vec b, \vec a\vec e)
    \end{equation}
\end{lemma}

\begin{proof}
    By induction on \(e\).
    \begin{itemize}
        \item If \(e = a \in \Act\), then \(\gamma(e) = \gamma(a) = (a, \checkmark)^\rho\).
        By reflexivity, \(\T^* \vdash e = a\).

        \item If \(e = \sigma(e_1, \dots, e_n)\), let \(\gamma(e_i) = t_i^\rho((\vec b_{i}, \checkmark), (\vec a_{i}, \vec e_{i}))\) for each \(i\).\footnote{
            Here, \(\vec a_i\) stands for a sequence of the form \(a_{1i}, a_{2i}, \dots a_{mi}\).
        }
        Then
        \[
            \gamma(e) = \sigma^\rho(
                t_1^\rho((\vec b_{1}, \checkmark), (\vec a_{1}, \vec e_{1}))
                , \dots,
                t_n^\rho((\vec b_{n}, \checkmark), (\vec a_{n}, \vec e_{n}))
            )
        \]
        and so we can prove the claim by deriving:
        \begin{align*}
            \T^* \vdash e
            &= \sigma(e_1, \dots, e_n) \\
            &= \sigma(
                t_1(\vec b_{1}, \vec a_{1} \vec e_{1}), \dots,
                t_n(\vec b_{n}, \vec a_{n} \vec e_{n})
            ) \tag{ind.~hyp.}
        \end{align*}

        \item If \(e = e_1 e_2\), let \(\gamma(e_1) = t_1^\rho((\vec b, \checkmark), (\vec a, \vec f))\).
        Then \(\gamma(e) = t_1^\rho((\vec b, e_2), (\vec a, \vec fe_2))\).
        By the induction hypothesis, \(\T^* \vdash e_1 = t_1(\vec b, \vec a\vec f)\), so we can derive:
        \[
            \T^* \vdash e_1e_2 = t_1(\vec b, \vec a\vec f)e_2 \stackrel{\text{(D)}}{=} t_1(\vec be_2, \vec a \vec f e_2)
        \]

        \item If \(e = e_1^{(s)}e_2\), let \(\gamma(e_1) = t_1^\rho((\vec b, \checkmark), (\vec a, \vec f))\) and let \(\gamma(e_2) = t_2^\rho((\vec d, \checkmark), (\vec c, \vec g))\).
        Then
        \begin{align*}
            \T^* \vdash e
            &= e_1^{(s)}e_2 \\
            &= s\Big(e_1(e_1^{(s)}e_2), e_2\Big) \tag{U}\\
            &= s\Big(t_1(\vec b, \vec a\vec f)(e_1^{(s)}e_2), e_2\Big) \tag{ind.~hyp.}\\
            &= s\Big(t_1(\vec b(e_1^{(s)}e_2), \vec a\vec f(e_1^{(s)}e_2)), e_2\Big) \tag{D}\\
            &= s\Big(t_1(\vec b(e_1^{(s)}e_2), \vec a\vec f(e_1^{(s)}e_2)), t_2(\vec d, \vec c\vec g)\Big) \tag{ind.~hyp.}
        \end{align*}
        This is the desired equivalence, because
        \begin{align*}
            \gamma(e_1^{(s)}e_2)
            &= s^\rho(t_1^\rho((\vec b, e_1^{(s)}e_2), (\vec a, \vec fe_1^{(s)}e_2)), \gamma(e_2)) \\
            &= s^\rho(t_1^\rho((\vec b, e_1^{(s)}e_2), (\vec a, \vec fe_1^{(s)}e_2)), t_2^\rho((\vec d, \checkmark), (\vec c, \vec g)))
        \end{align*}
    \end{itemize}
\end{proof}

\propSolutionsAreHomoms*

\begin{proof}
    As mentioned, the first statement is \cref{lem:soundness lemma}.
    For the second part, let us check that \(\phi \colon X \to \Exp\) is a solution if and only if \([-]_\equiv \circ \phi\) is a homomorphism of \(M\)-systems, i.e., $B_M([-]_\equiv \circ \phi) \circ \beta = [\gamma]_\equiv \circ [-]_\equiv \circ \phi$.

    First, we note the following: if $x \in X$ with \(\beta(x) = t_1^\rho((\vec b, \checkmark), (\vec a, \vec x))\), then:
    \begin{align*}
        B_M([-]_\equiv \circ \phi) \circ \beta(x)
        &= B_M([-]_\equiv) \circ B_M(\phi) \circ \beta(x) \tag{functoriality} \\
        &= B_M([-]_\equiv) \circ B_M(\phi) (t_1^\rho((\vec b, \checkmark), (\vec a, \vec x))) \tag{def.} \\
        &= B_M([-]_\equiv)(t_1^\rho((\vec b, \checkmark), (\vec a, \phi(\vec x)))) \tag{def.~\(B_M\)} \\
        &= t_1^\rho((\vec b, \checkmark), (\vec a, [\phi(\vec x)]_\equiv)) \tag{def.~\(B_M\)}
    \end{align*}
    (\(\Rightarrow\)) Assume that \(\phi\) is a solution, so that \(\T^* \vdash \phi(x) = t_1(\vec b, \vec a\phi(\vec x))\).
    By definition of \([\gamma]_\equiv\), we therefore have
    \begin{align*}
        ([\gamma]_\equiv \circ [-]_\equiv \circ \phi)(x)
        &= [\gamma]_\equiv([\phi(x)]_\equiv) \\
        &= [\gamma]_\equiv([t_1(\vec b, \vec a\phi(\vec x))]_\equiv) \\
        &= B_M([-]_\equiv)(\gamma(\phi(x))) \\
        &= t_1^\rho((\vec b, \checkmark), (\vec a, [\phi(\vec x)]_\equiv))
    \end{align*}
    By the derivation above, the last expression is equal to $B_M([-]_\equiv \circ \phi) \circ \beta(x)$.

    \noindent(\(\Leftarrow\)) Conversely, suppose $[-]_\equiv \circ \phi$ is a homomorphism.
    Now let \(\gamma(\phi(x)) = t_2^\rho((\vec d, \checkmark), (\vec c, \vec e))\).
    We then have \([\gamma]_\equiv \circ [-]_\equiv \circ \phi(x) = t_2^\rho((\vec d, \checkmark), (\vec c, [\vec e]_\equiv))\).
    Because \([\gamma]_\equiv \circ [-]_\equiv \circ \phi(x) = B_M([-]_\equiv \circ \phi) \circ \beta(x)\) and by the derivation above, we find
    \[
        t_1^\rho((\vec b, \checkmark), (\vec a, [\phi(\vec x)]_\equiv))
            = t_2^\rho((\vec d, \checkmark), (\vec c, [\vec e]_\equiv))
    \]
    which tells us by a standard property of free algebra constructions\footnote{Namely, that they are completely axiomatized by their corresponding theory.} that
    \begin{equation}
        \label{eq:square sides}
        \T \vdash t_1((\vec b, \checkmark), (\vec a, [\phi(\vec x)]_\equiv)) = t_2((\vec d, \checkmark), (\vec c, [\vec e]_\equiv))
    \end{equation}

    Then by any substitution where \((b_i, \checkmark) \mapsto b_i\), \((a_i, [\phi(x_i)]_\equiv) \mapsto a_i\phi(x_i)\), \((d_i, \checkmark) \mapsto d_i\), and \((c_i, [e_i]_\equiv) \mapsto c_i e_i\), the rule (\(\T\)) tells us that
    \[
        \T^* \vdash t_1(\vec b, \vec a\phi(\vec x)) = t_2(\vec d, \vec c\vec e)
    \]
    Note that the particular representatives of $[\phi(x_i)]_\equiv$ and $[e_i]_\equiv$ do not matter, because $\T^*$ includes the congruence rule.
    By \cref{lem:fundamental}, we can then conclude
    \[
        \T^* \vdash \phi(x)
            = t_2(\vec d, \vec c\vec e) = t_1(\vec b, \vec a\phi(\vec x))
    \]
\end{proof}

\propSolutionPullbacks*

\begin{proof}
    The first statement follows from \cref{prop:solutions are homoms}: \(\phi \circ h\) is a solution if and only if \([-]_\equiv \circ \phi \circ h\) is a homomorphism of \(M\)-systems.
    Since compositions of homomorphisms are homomorphisms, if \(\phi\) is a solution, then \([-]_\equiv \circ \phi\), and therefore \(([-]_\equiv \circ \phi) \circ h\) is a homomorphism, as desired.

    The second statement is a consequence of the definition of \(\langle e \rangle\) in \cref{prop:Exp locally finite}.
    Let \(\iota \colon \langle e \rangle \into \Exp\) be the inclusion map.
    Then \(\iota\) is a homomorphism of \(M\)-systems.
    Hence, \([-]_\equiv \circ \iota\) is a homomorphism of \(M\)-systems.
    This tells us that \(\iota\) is a solution to \(\langle e \rangle\).
\end{proof}

\begin{remark}
    It is worth noting here that the fact that \(\langle e \rangle \into \Exp\) is a solution could also be derived from \cref{lem:fundamental}.
\end{remark}

\section{Proofs of \cref{prop:Exp well-layered,lem:inverse image,thm:well-layered M-system covariety}}

We are going to show that the entry/body labelling of \((\Exp, \gamma)\) where the relation \(\eo\) is derived from the rules below is well-layered.
Recall that we write \(e \tr{a} \xi\) if \((a, \xi) \in \supp(\gamma(e))\) and \(e \to \xi\) if there exists an \(a \in \Act\) such that \(e \tr{a} \xi\).
\begin{gather}
    \label{eq:entry/body of Exp}
    \infer{e_1 \to \checkmark}{e_1 ^{(s)} e_2 \eo e_1 ^{(s)} e_2}
    \qquad
    \infer{e_1 \to f \quad f \to^+ \checkmark}{e_1 ^{(s)} e_2 \eo f(e_1 ^{(s)} e_2)}
    \qquad
    \infer{e_1 \eo f}{e_1e_2 \eo f e_2}
\end{gather}

Fix a free-algebra construction \((M, \eta, \rho)\) for \(\T\) for the signature \(S\).
The following technical lemma is convenient to have.

\begin{lemma}%
    \label{lem:support facts}
    Let \(\supp \colon M \Rightarrow \P\) be a support for \(\T\).
    \begin{enumerate}
        \item If \(t = t(x_1, \dots, x_n) \in S^*X\), then \(\supp_X(t^\rho) \subseteq \{x_1,\dots, x_n\}\).
        \item Let \(t_1 \in S^*X\).
        If \(x \in \supp(t_1^\rho)\), then for any \(t_2\) such that \(\T \models t_1 = t_2\), \(x\) appears in \(t_2\) as a variable.
    \end{enumerate}
\end{lemma}

\begin{proof}
    Let us start with the first claim.
    By induction on \(t\).
    If \(t = x \in X\), then by definition of \(\supp\), \(\supp_X(x^\rho) = \supp_X \circ \eta(x) = \{x\}\).
    If \(t = \sigma(t_1(\vec x), \dots, t_n(\vec x))\), then by definition of \(\supp\), \(\supp_X(t^\rho) \subseteq \bigcup_{i} \supp_X(t_i) \subseteq \{x_1, \dots, x_n\}\).

    On to the second claim.
    Let \(t_2 = t_2(x_1, \dots, x_n)\) such that every \(x_i\) appears in \(t_2\).
    By definition of \(\supp\), \(\supp_X(t_2^\rho) \subseteq \{x_1, \dots x_n\}\).
    Hence, if \(x \in \supp_X(t_1^\rho) = \supp_X(t_2^\rho)\), then \(x = x_i\) for some \(i\), which means that $x$ appears in $t_2$.
\end{proof}

A path of the form \(x_1 \to x_2 \to \dots \to x_n \to x_1\) is called a \emph{cycle}.
A \emph{simple cycle} is a path of the same form with \(x_i \neq x_j\) for \(i \neq j\).
It is worth noting that (in an arbitrary directed graph), every cycle is the concatenation of finitely many simple cycles.
We need the following characterization of loops in $(\Exp, \gamma)$.

\begin{lemma}%
    \label{lem:cycle form}
    Every simple cycle in \((\Exp, \to)\) is of the form
    \begin{equation}
        \label{eq:cycle form}
        \begin{aligned}
            &(e_1^{(s)} e_2)g_1\dots g_k
            \to f_1 (e_1^{(s)} e_2)g_1\dots g_k
            \to \dots\\
            &\hspace{8em}\dots \to f_n (e_1^{(s)} e_2)g_1\dots g_k \to (e_1^{(s)} e_2)g_1\dots g_k
        \end{aligned}
    \end{equation}
    where \(e_1 \to f_1 \to \dots \to f_n \to \checkmark\) and the arrangement of parentheses in \(g_1\cdots g_k\) is arbitrary.
\end{lemma}

\begin{proof}
    By induction on \(h \in \Exp\), we will show that if \(h \to^* e\) and \(e\) is contained in a simple cycle, then that simple cycle is of the form~\eqref{eq:cycle form}.
    \begin{itemize}
        \item In case \(h = a \in \Act\), the statement of the lemma is vacuous because \(\langle a\rangle\) does not contain any cycle.

        \item Let \(h = \sigma(h_1, \dots, h_n)\).
        If \(\sigma(h_1, \dots, h_n) \tr{a} f\), then we know that
        \begin{align*}
            (a, f)
                &\in \supp(\gamma(\sigma(h_1, \dots, h_n))) \\
                &= \supp(\sigma^\rho(\gamma(h_1), \dots, \gamma(h_n))) \\
                &\subseteq \bigcup\nolimits_i \supp(\gamma(h_i))
        \end{align*}
        so \(h_i \to f\) for some \(i\).
        The rest follows from the induction hypothesis.

        \item Let \(h = h_1h_2\) and suppose \(h \to^* e\) where \(e\) appears in a simple cycle.
        If \((a, e) \in \supp(\gamma(h))\), then \((a, e)\) appears in every term \(t((\vec b, h_2), (\vec a, \vec eh_2))\) such that \(\gamma(h) = t^\rho\), by \Cref{lem:support facts}.
        Then \(e\) is of the form \(h_1'h_2\) with \(h_1 \to h_1'\) or \(e = h_2\).
        So, by induction on the length of the path \(h \to \cdots \to e\), either \(e\) is of the form \(h_1'h_2\) with \(h_1 \to^+ h_1'\) or \(h_2 \to^* e\).
        If \(h_2 \to^* e\), then the rest follows from the induction hypothesis.
        So, suppose \(e = h_1'h_2\) and that we don't have \(h_2 \to^+ e\).
        If \(h_1 \to^+ h_1'\) and \(e = h_1'h_2\) is in a simple cycle, then there must exist a cycle of the form \(
            h_1' \to k_1 \to \cdots \to k_m \to h_1'
        \).
        By the induction hypothesis, since \(h_1 \to^* h_1'\), this simple cycle is of the form~\eqref{eq:cycle form}.
        Without loss of generality, we can assume that we have \(h_1' = (e_1^{(s)}e_2)g_1\cdots g_k\), \(k_i = f_i(e_1^{(s)}e_2)g_1\cdots g_k\) for \(e_1 \to f_1 \to \cdots \to \checkmark\).
        Then we have \(h \to^* h_1'h_2 = (e_1^{(s)}e_2)g_1\cdots g_k h_2 \to f_1(e_1^{(s)}e_2)g_1\cdots g_k h_2 \to \cdots \to (e_1^{(s)}e_2)g_1\cdots g_k h_2\), which is in the form of~\eqref{eq:cycle form}.

        \item Now let \(h = h_1^{(r)}h_2\), and suppose that \(h \to^* e\) such that \(e\) appears in a simple cycle.
        If \((a, e) \in \supp(\gamma(h))\), then \((a, e)\) appears in every term \(s(t((\vec b, h_1^{(r)}h_2), (\vec a, \vec e(h_1^{(r)}h_2))),h_2)\) where \(\gamma(h) = s^\rho(t^\rho, \gamma(h_2))\), by \Cref{lem:support facts}.
        Then \(e\) is of the form \(h_1'(h_1^{(r)}h_2)\) with \(h_1 \to h_1'\) or \(h_2 \to e\).
        By induction on the length of the path \(h \to \cdots \to e\), we can see that \(e\) is either of the form \(h_1'(h_1^{(r)}h_2)\) with \(h_1 \to^+ h_1'\) or \(h_2 \to^+ e\).
        In the latter case, the rest follows from the induction hypothesis.
        In the former case, there are two possibilities:
        \begin{enumerate}
            \item We could have \(e = h_1'(h_1^{(r)}h_2) \to \cdots \to h_1^{(s)}h_2 \to \cdots \to h_1'(h_1^{(r)}h_2)\) where \(h_1' \to^+ \checkmark\).
            In this case, \(e\) is contained in a simple cycle of the form~\eqref{eq:cycle form} by taking \(k = 0\), \(e_1 = h_1\), \(s = r\), and \(e_2 = h_2\).

            \item Or, we could have \(e = h_1'(h_1^{(r)}h_2) \to \cdots \to h_1'(h_1^{(r)}h_2)\) avoid \(h_1^{(r)}h_2\) altogether.
            In this case, we must have \(h_1' \to k_1 \to \cdots \to k_m \to h_1'\).
            By the induction hypothesis applied to \(h_1 \to^* h_1'\), without loss of generality we can assume that this cycle is of the form \(h_1' = (e_1^{(s)} e_2)g_1\dots g_k\) and \(k_i = f_i(e_1^{(s)} e_2)g_1\dots g_k\) for \(f_1 \to f_2 \to \cdots \to \checkmark\).
            In this case, \(e = h_1'(h_1^{(r)}h_2) = (e_1^{(s)} e_2)g_1\dots g_k(h_1^{(r)}h_2) \to f_1(e_1^{(s)} e_2)g_1\dots g_k(h_1^{(r)}h_2) \to \cdots \to e\) as desired.
        \end{enumerate}
    \end{itemize}
\end{proof}

\propExpWellLayered*

For the entry/body layering above, we need to show that
\begin{enumerate}
    \item We do not have \(e \bo^+ e\) for any \(e \in \Exp\).
    \item For any \(e,f \in \Exp\), if \(e \eo f\), then \(f \bo^+ e\).
    \item The directed graph \((\Exp, \diredge)\) is acyclic.
    \item For any \(e,f \in \Exp\), if \(e \diredge f\), then we do not have \(f \to \checkmark\).
\end{enumerate}

\begin{proof}
    We already saw that \((\Exp,\gamma)\) is locally finite in \cref{prop:Exp locally finite}.
    Let us check that it satisfies the other properties.

    Suppose \(e \to \cdots \to e\).
    We can assume without loss of generality that this is a simple cycle, since every cycle is a concatenation of simple cycles.
    Then by \cref{lem:cycle form}, \(e \to \cdots \to e\) is of the form~\eqref{eq:cycle form}.
    By definition of the entry/body labelling, the first transition in~\eqref{eq:cycle form} is a loop entry transition, \((e_1^{(s)} e_2)g_1\dots g_k \eo f_1 (e_1^{(s)} e_2)g_1\dots g_k\).
    Hence, at least one transition in \(e \to \cdot \to e\) is an \(\eo\) transition.
    This shows property 1.

    To see property 2, observe that by definition of the entry/body labelling, every loop entry transition \(e \eo e'\) such that \(e' \neq e\) is of the form \((e_1^{(s)} e_2)g_1\dots g_k \eo f_1 (e_1^{(s)} e_2)g_1\dots g_k\) where \(f_1 \to^+ \checkmark\).
    Since \(f_1 \to^+ \checkmark\), \(f_1 (e_1^{(s)} e_2)g_1\dots g_k \to^+ (e_1^{(s)} e_2)g_1\dots g_k\).
    This establishes property 2.

    To see property 3, we follow Grabmayer and Fokkink and add extra data to loop entry and body transitions, specifically a number \(\tr{[n]}\) to transitions such that \(\bo\) transitions are those with \(n = 0\) and \(\eo\) transitions with \(n > 0\).
    Define the \emph{star height} of an expression \(|e|_*\) by setting
    \begin{mathpar}
        |a|_* = 0
        \and
        |\sigma(e_1, \dots, e_n)|_* = \max_i |e_i|_*
        \\
        |e_1 e_2|_* = \max_i |e_i|_*
        \and
        |e_1^{(s)}e_2|_* = \max \{|e_1|_* + 1, |e_2|\}
    \end{mathpar}
    We define \(e \tr{[n]} f\) such that the label is given by the smallest \(n \in \N\) for which the following rules are satisfied.
    \begin{gather}
        \label{eq:entry/body of Exp numbered}
        \infer{e_1 \tr{[n]} e_1'}{e_1e_2 \tr{[n]} e_1'e_2}
        \qquad
        \infer{e_1 \to \checkmark}{e_1^{(s)}e_2 \tr{[|e_1|_* + 1]} e_1^{(s)}e_2}
        \qquad
        \infer{e_1 \to e_1' \quad e_1' \to^+ \checkmark}{e_1^{(s)}e_2 \tr{[|e_1|_* + 1]} e_1'(e_1^{(s)}e_2)}
    \end{gather}
    Comparing with \cref{eq:entry/body of Exp}, we see that \(e \tr{[n]} f\) for \(n > 0\) if and only if \(e \eo f\).

    We will now show that if \(e_1 \eo f_1 \bo \cdots \bo f_n \bo e_2 \eo g\) and \(e_1 \notin \{f_1, \dots, f_n, e_2, g\}\), and if \(e_1 \tr{[n]} f_1\) and \(e_2 \tr{[m]} g\), then \(m < n\).
    To this end, we begin with the following observations about star height:
    \begin{enumerate}
        \item[(A)] \(|e|_* < |e^{(s)}f|_*\), and
        \item[(B)] if \(e \to f\), then \(|e|_* \ge |f|_*\).
    \end{enumerate}
    Observation (A) is by definition.
    Observation (B) follows from an easy induction on \(e\).
    It follows from (A) and (B) that if \(e \to^+ e'\), then \(|e'|_* < |e^{(s)}f|_*\).

    Now, every transition \(e_1 \eo f_1\) is of the form \(e_1 = (h^{(s)}k)\ell_1\cdots \ell_m \eo h_1(h^{(s)}k)\ell_1\cdots \ell_m\) such that \(h \to h_1 \to^+ \checkmark\).
    This allows us to rewrite the path \(e_1 \eo f_1 \bo \cdots \bo f_n \bo e_2 \eo g\) as
    \begin{align*}
        e_1 &=(h^{(s)}k)\ell_1\cdots \ell_m
        \eo h_1(h^{(s)}k)\ell_1\cdots \ell_m
        \bo h_2(h^{(s)}k)\ell_1\cdots \ell_m
        \bo \dots \\
        &\hspace{8em}\dots \bo h_n(h^{(s)}k)\ell_1\cdots \ell_m
        = e_2 \eo g
    \end{align*}
    where \(h \to h_1 \to \cdots \to h_n\), because \(e_1\) does not appear twice in this path.
    Now, by definition of \(\tr{[n]}\), \[
        (h^{(s)}k)\ell_1\cdots \ell_m \etr{[|h|_* + 1]} h_1(h^{(s)}k)\ell_1\cdots \ell_m
    \]
    The second \(\eo\) transition must therefore have \(h_n = (p^{(s)}q)u_1\dots u_i\) and
    \[
        (p^{(s)}q) u_1\dots u_i (h^{(s)}k) \ell_1\cdots \ell_m
        \eo p'(p^{(s)}q) (h^{(s)}k)\ell_1\cdots \ell_m = g
    \]
    with \(p' \to^+ \checkmark\), because it is a loop entry transition.
    But here, by definition of \(\tr{[n]}\) again,
    \[
        (p^{(s)}q) u_1\dots u_i (h^{(s)}k) \ell_1\cdots \ell_m
        \etr{[|p|_* + 1]} p'(p^{(s)}q) (h^{(s)}k)\ell_1\cdots \ell_m = g
    \]
    Using the observations again,
    \begin{align*}
        |p|_* + 1
        &\le |h_n|_* \tag{\(h_n = (p^{(s)}q)u_1\cdots u_i\)} \\
        &\le |h|_* \tag{\(h \to^+ h_n\)}\\
        &< |h|_* + 1
    \end{align*}
    This concludes the proof of property 3.

    Finally, we turn to property 4.
    Let \(e_1 \eo f_1 \bo \cdots \to f_n \bo e_2\) with \(e_1 \notin \{f_1, \dots, f_n, e_2\}\).
    As before, every transition of the form \(e_1 \eo f_1\) is of the form \((h ^{(s)} k)g_1\cdots g_m \eo h_1(h ^{(s)} k)g_1\cdots g_m\) with \(h \to h_1 \to^+ \checkmark\).
    Now, since \(e_1 \neq f_i\) and \(e_1 \neq e_2\), we also know that \(f_i = h_i(h ^{(s)} k)g_1\cdots g_m\) and \(e_2 = f(h ^{(s)} k)g_1\cdots g_m\) with \(h_1 \to \cdots \to h_n \to f\).
    There are no transitions \(ef \to \checkmark\) for any \(e,f \in \Exp\), so in particular, \(e_2 = f(h ^{(s)} k)g_1\cdots g_m \not\to \checkmark\).
\end{proof}

\section{Proofs of \cref{prop:canonical solution unique} and Completeness (\cref{thm:completeness})}

We already noted the following property informally in the main text.

\begin{lemma}%
	\label{lem:star dist}
	For \(e,f,g\in\Exp\) and $s(u,v) \in S^*\Var$, \(\T^* \vdash (e^{(s)}f)g = e^{(s)}(fg)\).
\end{lemma}

\begin{proof}
    We simply derive as follows:
	\begin{align*}
		\T^* \vdash (e^{(s)}f)g
		&= s(e(e^{(s)}f), f)g \tag{U} \\
		&= s(e(e^{(s)}f)g, fg)\tag{D} \\
		&= e^{(s)}(fg) \tag{RSP}
	\end{align*}
\end{proof}

The following properties are crucial in proving \cref{prop:canonical solution unique}.

\begin{lemma}%
	\label{lem:intermediate solution}
	Let \((X, \beta)\) be a well-layered \(M\)-system, with entry/body labeling $\eo, \bo$, and $\phi_\beta$, $\tau_\beta$ and $|{-}|_{bo}$ and $|{-}|_{en}$ as in \Cref{def:system canonical solution}.
    If \(x \diredge y\) in \((X, \beta)\), then the following hold:
	\begin{enumerate}
		\item \(\T^* \vdash \phi_\beta(y) = \tau_\beta(y, x)\phi_\beta(x)\), and
		\item for any solution \(\varphi\) to \((X, \beta)\), \(\T^* \vdash \varphi(y) = \tau_\beta(y, x)\varphi(x)\).
	\end{enumerate}
\end{lemma}

\begin{proof}
	We prove the first claim by induction on \(|y|_{bo}\).
	Suppose \(x \diredge y\), and let
    \begin{equation}%
        \label{eq:factorize-beta-y}
        \beta(y) = s^\rho\Big( t_1^\rho((\vec a, y), (\vec b, \vec x)),~ t_2^\rho((\vec c, x) , (\vec d, \vec z))\Big) 
    \end{equation}
    where $\vec{x}$ is a vector such that \(y \neq x_i\) and \(y \eo x_i\) for each \(i\), and $\vec{z}$ is a vector such that \(x \neq z_k\) and \(y \bo z_k\) for each \(k\).
    Note in particular that if $y \to x$, then $y \bo x$, because if $y \eo x$ then $y \diredge x$, which would create a cycle in $(X, \diredge)$.
    Furthermore, this factorisation of $\beta(y)$ is possible not just because $\T$ is malleable, but also because $y \not\to \checkmark$, i.e., $(a', \checkmark) \notin \supp_X(\beta(y))$ for any \(a'\).

    With this in mind, we can calculate that
    \begin{align}
        \phi_\beta(y)
        &= r^{(s)} \Big( t_2(\vec{c}\phi_\beta(x), d_1\phi_\beta(z_1), \dots, d_m\phi_\beta(z_m)) \Big) \tag*{} \\
        \label{eq:tau-beta-y-x}
        \tau_\beta(y, x)
        &= r^{(s)} \Big(t_2(\vec c, d_1\tau_\beta(y_1, x), \dots, d_m\tau_\beta(y_m, x))
        \Big)
    \end{align}
    with $r = t_1(\vec a, b_1\tau_\beta(x_1, y), \dots, b_n\tau_\beta(x_n, y))$.
    We can now derive as follows:
	\begin{align*}
		\T^*\vdash  {}
        &\tau_\beta(y, x) \phi_\beta(x) \\
		&= r^{(s)}
            \Big(
                t_2(\vec c, d_1\tau_\beta(y_1, x), \dots, d_m\tau_\beta(y_m, x))
            \Big)
            \phi_\beta(x)
            \tag{def. $\tau_\beta(y,x)$} \\
		&= r^{(s)}
            \Big(
                t_2(\vec c, d_1\tau_\beta(y_1, x), \dots, d_m\tau_\beta(y_m, x))
                \phi_\beta(x)
            \Big)
            \tag{\cref{lem:star dist}} \\
		&= r^{(s)}
            \Big(
                t_2(\vec c \phi_\beta(x), d_1\tau_\beta(y_1, x)\phi_\beta(x), \dots, d_m\tau_\beta(y_m, x)\phi_\beta(x))
            \Big)
            \tag{D} \\
		&= r^{(s)}
            \Big(
                t_2(\vec c \phi_\beta(x), d_1\phi_\beta(y_1), \dots, d_m\phi_\beta(y_m))
            \Big)
            \tag{IH, \(|y_i|_{bo} < |y|_{bo}\)} \\
		&= \phi_\beta(y)
            \tag{def. $\phi_\beta(x)$}
	\end{align*}

	For the second claim, we proceed by induction on \((|x|_{eo}, |y|_{bo})\) in the lexicographical ordering.
	Assume \(\varphi\) is a solution to \((X, \beta)\) and \(x \diredge y\).
	Let $\beta(y)$ be factorized as in~\eqref{eq:factorize-beta-y}, which gives rise to the form of $\tau_\beta(y, z)$ in~\eqref{eq:tau-beta-y-x}.
    We derive:
	\begin{align*}
		\T^*\vdash
		\varphi(y)
		&= s\Big(t_1(\vec a\varphi(y), \vec b\varphi(\vec x)), t_2(\vec c\varphi(x), \vec d\varphi(\vec y))\Big) \tag{$\varphi$ is a solution} \\
		&= s\Big(t(\vec a\varphi(y), \vec b\varphi(\vec x)), t_2(\vec c\varphi(x), \vec d\tau_\beta(\vec y, x)\varphi(x))\Big)
		\tag{IH, \(|y_i|_{bo} < |y|_{bo}\)}\\
		&= s\Big(t_1(\vec a\varphi(y), \vec b\varphi(\vec x)), t_2(\vec c, \vec d\tau_\beta(\vec y, x))\varphi(x)\Big)
		\tag{D}\\
		&= s\Big(t_1(\vec a\varphi(y), \vec b\tau_\beta(\vec x, y)\varphi(y)), t_2(\vec c, \vec d\tau_\beta(\vec y, x))\varphi(x)\Big)
		\tag{IH, \((|y|_{eo}, |x_i|_{bo}) < (|x|_{eo}, |y|_{bo})\)}\\
		&= s\Big(t_1(\vec a, \vec b\tau_\beta(\vec x, y))\varphi(y), t_2(\vec c, \vec d\tau_\beta(\vec y, x))\varphi(x)\Big)
		\tag{D}\\
		&= t_1(\vec a, \vec b\tau_\beta(\vec x, y))^{(s)} (t_2(\vec c, \vec d\tau_\beta(\vec y, x))\varphi(x))
		\tag{RSP}\\
		&= \Big(t_1(\vec a, \vec b\tau_\beta(\vec x, y))^{(s)} t_2(\vec c, \vec d\tau_\beta(\vec y, x))\Big)\varphi(x)
		\tag{\cref{lem:star dist}}\\
		&= \tau_\beta(y, x)\varphi(x) \tag{def. $\tau_\beta(y, x)$}
	\end{align*}
\end{proof}

\propCanonicalSolutionUnique*
\begin{proof}
	Let us begin by showing that it is indeed a solution.
    First, write
	\[
        \beta(x) = \mathcolor{red}{s^\rho\Big(} t_1^\rho((\vec a, x), (\vec b, \vec x))\mathcolor{red}{,}t_2^\rho((\vec c, \checkmark) , (\vec d, \vec y))\mathcolor{red}{\Big)} 
	\]
    where $\vec{x}$ is a vector such that \(x \neq x_i\) and \(x \eo x_i\) for each \(i\), and $\vec{y}$ is a vector such that \(x \bo y_j\) for each \(j\).
	Then by definition,
	\begin{align*}
		\T^* \vdash \phi_\beta(x)
		&= \Big(t_1(\vec a, \vec b\tau_\beta(\vec x, x))\Big)^{(s)}\Big( t_2(\vec c, \vec d\phi_\beta(\vec y))\Big) \\
		&= s\Big(t_1(\vec a, \vec b\tau_\beta(\vec x, x))\phi_\beta(x), t_2(\vec c, \vec d\phi_\beta(\vec y))\Big) \tag{U}\\
		&= s\Big(t_1(\vec a\phi_\beta(x), \vec b\tau_\beta(\vec x, x)\phi_\beta(x)), t_2(\vec c, \vec d\phi_\beta(\vec y))\Big) \tag{D}\\
		&= s\Big(t_1(\vec a\phi_\beta(x), \vec b\phi_\beta(\vec x)), t_2(\vec c, \vec d\phi_\beta(\vec y))\Big) \tag{\cref{lem:intermediate solution}}
	\end{align*}
	as desired.
	To see that \(\phi_\beta\) is the unique solution, we show that for any solution \(\varphi\), \(\T^* \vdash \varphi(x) = \phi_\beta(x)\) by induction on \(|x|_{bo}\), as follows.
	\begin{align*}
		\T^* \vdash \varphi(x)
		&= s\Big(t_1(\vec a\varphi(x), \vec b\varphi(\vec x)), t_2(\vec c, \varphi(\vec y))\Big) \\
		&= s\Big(t_1(\vec a\varphi(x), \vec b\varphi(\vec x)), t_2(\vec c, \phi_\beta(\vec y))\Big) \tag{IH, \(|y_i|_{bo} < |x|_{bo}\)} \\
		&= s\Big(t_1(\vec a\varphi(x), \vec b\tau_\beta(\vec x, x)\varphi(x)), t_2(\vec c, \phi_\beta(\vec y))\Big) \tag{\cref{lem:intermediate solution}} \\
		&= s\Big(t_1(\vec a, \vec b\tau_\beta(\vec x, x))\varphi(x), t_2(\vec c, \phi_\beta(\vec y))\Big) \tag{D}\\
		&= \Big(t_1(\vec a, \vec b\tau_\beta(\vec x, x))\Big)^{(s)}\Big( t_2(\vec c, \phi_\beta(\vec y))\Big) \tag{RSP} \\
		&= \phi_\beta(x)
	\end{align*}
\end{proof}

\thmCompleteness*

\begin{proof}
Because $e_1 \bisim e_2$, they are also bisimilar as states of $\langle e_1 \rangle$ and $\langle e_2 \rangle$ respectively.
Then there is an $M$-system \((Z, \delta)\) and homomorphisms of \(M\)-systems \(h_i \colon \langle e_1 \rangle \to (Z, \delta)\) for $i \in \{1, 2\}$ such that \(h_1(e_1) = h_2(e_2)\).
Let \(z = h_1(e_1) = h_2(e_2)\), and note that \(\langle z\rangle = h_1(\langle e_1\rangle) = h_2(\langle e_2 \rangle)\).
In particular, \(\langle z\rangle\) is the homomorphic image of \(\langle e_1\rangle\).
By \cref{prop:Exp well-layered}, \(\langle e_1\rangle\) and \(\langle e_2 \rangle\) are well-layered.
By \cref{thm:well-layered M-system covariety}, \(\langle z\rangle\) is also well-layered, because it is the homomorphic image of \(\langle e_1\rangle\).
By \cref{prop:canonical solution unique}, \(\langle e_1\rangle\), \(\langle e_2\rangle\), and \(\langle z\rangle\) all admit unique solutions.
Let \(\phi \colon \langle z \rangle \to \Exp\) be the solution to \(\langle z \rangle\).
By \cref{prop:solution pullbacks}, \(\phi \circ h_i\) is a solution to $\langle e_i \rangle$ for $i \in \{1,2\}$.
Since the inclusions \(\langle e_i \rangle \into \Exp\) are also solutions by \Cref{prop:solution pullbacks}, by uniqueness
\(
    \T^* \vdash e_1 = \phi \circ h_1(e_1) = \phi \circ h_2(e_2) = e_2
\), as desired.
\end{proof}

\section{Details of Examples and Nonexamples}

Let us verify that the few instantiations we described in the paper are indeed instantiations.

\begin{lemma}
    The equational theories \(\SL\) (Inst.~\ref{inst:semilattices}), \(\GA\) (Inst.~\ref{inst:guarded alg}), \(\CA\) (Inst.~\ref{inst:convex alg}), \(\GC\) (Inst.~\ref{inst:probgkat}), and \(\CS\) (convex semilattices) are all supported.
\end{lemma}

\begin{proof}
    The free-algebra construction for \(\SL\) is given by \((\P, \{-\}, \rho)\), where \({+^\rho} = {\cup}\) and \(0^\rho = \emptyset\).
    Define \(\supp = \id \colon \P \Rightarrow \P\).
    Clearly, \(\supp \circ \{-\} = \id \circ \{-\} = \{-\}\).
    Furthermore, given a term \(t\) in which the variables \(x_1, \dots, x_n\) appear \(t^\rho = \{x_1, \dots, x_n\}\).
    Hence, \(\supp(t^\rho) = t^\rho = \{x_1, \dots, x_n\}\).

    The free-algebra construction for \(\GA\) is given by \((\R,\eta, \rho)\), where \(\R = (\bot + (-))^\At\), \(\eta_X(x)(\alpha) = x\), \((\theta_1 +_b^\rho \theta_2)(\alpha) = \textbf{if}~\alpha \le b~\textbf{then}~\theta_1(\alpha)~\textbf{else}~\theta_2(\alpha)\), and \(0^\rho(\alpha) = \bot\).
    Define \(\supp_X(\theta) = \theta(\At)\setminus\{\bot\}\).
    It is straightforward to see that this is natural.
    We furthermore have \(\supp \circ \eta_X(x) = \eta_X(x)(\At) = \{x\}\).
    To see that for any term \(t = t(\vec x)\), \(\supp(t^\rho) \subseteq \{x_1, \dots, x_n\}\), we proceed by induction on \(t\).
    \begin{itemize}
        \item If \(t = 0\), then \(\supp_X(t^\rho) = \emptyset\).
        \item If \(t = x\), then \(\supp_X(t^\rho) = \supp_X(\eta_X(x)) = \{x\}\).
        \item If \(t = t_1(\vec x) +_b t_2(\vec y)\), then
        \begin{align*}
            \supp_X(t^\rho)
            &= t^\rho(\At)\setminus \{\bot\} \\
            &= (t_1^\rho(\{\alpha \le b\}) \cup t_2^\rho(\{\alpha \le \bar b\}))\setminus \{\bot\} \\
            &\subseteq (t_1^\rho(\At) \cup t_2^\rho(\At))\setminus \{\bot\} \\
            &= \supp_X(t_1^\rho) \cup \supp_X(t_2^\rho) \\
            &\subseteq \{\vec x\} \cup \{\vec y\} \tag{IH}
        \end{align*}
    \end{itemize}

    The free-algebra construction for \(\CA\) is given by \((\D,\eta, \rho)\)~\cite{swirszcz-74}, where \(\D\) is the finitely supported probability distribution functor, \(\eta_X(x) = \delta_x\), \((\theta_1 \oplus_p^\rho \theta_2)(\alpha) = p\theta_1(\alpha) + (1-p)\theta_2\), and \(0^\rho = \delta_\bot\).
    Define \(\supp_X(\theta) = \{x \in X \mid \theta(x) > 0\}\).
    This is natural in $X$, because if $f: X \to Y$, then
    \begin{align*}
        \supp_Y(\D(\bot + f)(\theta))
            &= \supp_Y(\lambda y.\sum \{ \theta(x) \mid f(x) = y \}) \\
            &= \{ y \in Y \mid \exists x \in X.\ f(x) = y \wedge \theta(x) > 0 \} \\
            &= \{ f(x) \mid x \in X \wedge \theta(x) > 0 \} \\
            &= \P(f)(\supp_X(\theta))
    \end{align*}
    We furthermore have \(\supp \circ \eta_X(x) = \{y \in X \mid \delta_x(y) > 0\} = \{x\}\).
    To see that for any term \(t = t(\vec x)\), \(\supp(t^\rho) \subseteq \{x_1, \dots, x_n\}\), we proceed by induction on \(t\).
    \begin{itemize}
        \item If \(t = 0\), then \(\supp_X(t^\rho) = \{x \in X \mid \delta_\bot(x) > 0\} = \emptyset\).
        \item If \(t = x\), then \(\supp_X(t^\rho) = \supp_X(\delta_x) = \{x\}\).
        \item If \(t = t_1(\vec x) \oplus_p t_2(\vec y)\), then
        \begin{align*}
            \supp_X(t^\rho)
            &= \{x \in X \mid t^\rho(x) > 0\} \\
            &= \{x \in X \mid (pt_1^\rho + (1-p)t_2^\rho)(x) > 0\} \\
            &\subseteq \{x \in X \mid t_1^\rho(x) > 0\} \cup \{x \in X \mid t_2^\rho(x) > 0\} \\
            &= \supp_X(t_1^\rho) \cup \supp_X(t_2^\rho) \\
            &\subseteq \{\vec x\} \cup \{\vec y\} \tag{IH}
        \end{align*}
    \end{itemize}

    The free-algebra construction for \(\GC\) is given by \((\D(\bot + (-))^\At,\eta, \rho)\), where \(\eta_X(x)(\alpha) = \delta_x\), \((\chi_1 \oplus_p^\rho \chi_2)(\alpha) = p\chi_1(\alpha) + (1-p)\chi_2\), and \(0^\rho(\alpha) = \delta_\bot\).
    Define \(\supp_X(\chi) = \bigcup_{\alpha \in \At}\{x \in X \mid \chi(\alpha)(x) > 0\}\).
    Again, it is straightforward to check that \(\supp\) is natural.
    We furthermore have \(\supp \circ \eta_X(x) = \{y \in X \mid \delta_x(y) > 0\} = \{x\}\).
    To see that for any term \(t = t(\vec x)\), \(\supp(t^\rho) \subseteq \{x_1, \dots, x_n\}\), we proceed by induction on \(t\).
    \begin{itemize}
        \item If \(t = 0\), then \(\supp_X(t^\rho) = \supp_X(\alpha \mapsto \delta_\bot) = \{x \in X \mid \delta_\bot(x) > 0\} = \emptyset\).
        \item If \(t = x\), then \(\supp_X(t^\rho) = \supp_X(\alpha \mapsto \delta_x) = \{x\}\).
        \item If \(t = t_1(\vec x) \oplus_p t_2(\vec y)\), then
        \begin{align*}
            \supp_X(t^\rho)
            &= \bigcup_{\alpha \in \At}\{x \in X \mid t^\rho(\alpha)(x) > 0\} \\
            &= \bigcup_{\alpha \in \At}\{x \in X \mid (pt_1^\rho(\alpha) + (1-p)t_2^\rho(\alpha))(x) > 0\} \\
            &\subseteq \bigcup_{\alpha \in \At}\{x \in X \mid t_1^\rho(\alpha)(x) > 0\} \cup \bigcup_{\alpha \in \At}\{x \in X \mid t_2^\rho(\alpha)(x) > 0\} \\
            &= \supp_X(t_1^\rho) \cup \supp_X(t_2^\rho) \\
            &\subseteq \{\vec x\} \cup \{\vec y\} \tag{IH}
        \end{align*}
        \item If \(t = t_1(\vec x) +_b t_2(\vec y)\), then
        \begin{align*}
            \supp_X(t^\rho)
            &= \bigcup_{\alpha \in \At}\{x \in X \mid t^\rho(\alpha)(x) > 0\} \\
            &= \bigcup_{\alpha \le b}\{x \in X \mid t_1^\rho(\alpha)(x) > 0\} \cup \bigcup_{\alpha \le \bar b}\{x \in X \mid t_2^\rho(\alpha)(x) > 0\} \\
            &\subseteq \bigcup_{\alpha \in \At}\{x \in X \mid t_1^\rho(\alpha)(x) > 0\} \cup \bigcup_{\alpha \in \At}\{x \in X \mid t_2^\rho(\alpha)(x) > 0\} \\
            &= \supp_X(t_1^\rho) \cup \supp_X(t_2^\rho) \\
            &\subseteq \{\vec x\} \cup \{\vec y\} \tag{IH}
        \end{align*}
    \end{itemize}

    The free-algebra construction for \(\CS\) is given by \((\mathcal C, \eta, \rho)\), where \(\mathcal C X\) is the set of convex subsets of \(\D(\bot + X)\) containing \(\bot\), \(\eta_X(x) = \{p \delta_\bot + (1-p)\delta_x \mid p \in [0,1]\}\), \((U \oplus_p^\rho V) = \{p \theta_1 + (1-p)\theta_2 \mid p \in [0,1], \theta_1 \in U, \theta_2 \in V\}\), and \(U +^\rho V = \conv(U \cup V)\)~\cite{bonchi-sokolova-vignudelli-21,schmid-thesis}.
    Define \(\supp_X(U) = \bigcup_{\theta \in U}\{x \in X \mid \theta(x) > 0\}\).
    Again, it is straightforward to check that \(\supp\) is natural.
    We furthermore have \(\supp \circ \eta_X(x) = \{y \in X \mid \delta_x(y) > 0\} = \{x\}\).
    To see that for any term \(t = t(\vec x)\), \(\supp(t^\rho) \subseteq \{x_1, \dots, x_n\}\), we proceed by induction on \(t\).
    \begin{itemize}
        \item If \(t = 0\), then \(\supp_X(t^\rho) = \supp_X(\{\delta_\bot\}) = \{x \in X \mid \delta_\bot(x) > 0\} = \emptyset\).
        \item If \(t = x\), then \(\supp_X(t^\rho) = \supp_X(\{p \delta_\bot + (1-p)\delta_x \mid p \in [0,1]\}) = \{x\}\).
        \item If \(t = t_1(\vec x) +^\rho t_2(\vec y)\), then set \(U = t_1^\rho\) and \(V = t_2^\rho\) and calculate
        \begin{align*}
            \supp_X(t^\rho)
            &= \bigcup_{\theta \in \conv(U \cup V)} \{x \in X \mid \theta(x) > 0\} \\
            &= \bigcup_{p \in [0,1]}\bigcup_{\theta_1 \in U}\bigcup_{\theta_2 \in V} \{x \in X \mid (p \theta_1 + (1-p)\theta_2)(x) > 0\} \\
            &\subseteq \bigcup_{p \in [0,1]}\bigcup_{\theta_1 \in U}\bigcup_{\theta_2 \in V} (\{x \in X \mid \theta_1(x) > 0\} \cup \{x \in X \mid \theta_2(x) > 0\}) \\
            &= \bigcup_{\theta_1 \in U}\bigcup_{\theta_2 \in V} (\{x \in X \mid \theta_1(x) > 0\} \cup \{x \in X \mid \theta_2(x) > 0\}) \\
            &= \supp_X(U) \cup \supp_X(V) \\
            &\subseteq \{\vec x\} \cup \{\vec y\} \tag{IH}
        \end{align*}
        \item If \(t = t_1(\vec x) \oplus_p t_2(\vec y)\), then set \(U = t_1^\rho\) and \(V = t_2^\rho\) and calculate
        \begin{align*}
            \supp_X(t^\rho)
            &= \bigcup_{\theta_1 \in U} \bigcup_{\theta_2 \in V} \{x \in X \mid (p\theta_1 + (1-p)\theta_2)(x) > 0\} \\
            &\subseteq \bigcup_{\theta_1 \in U} \bigcup_{\theta_2 \in V} (\{x \in X \mid \theta_1(x) > 0\} \cup \{x \in X \mid \theta_2(x) > 0\}) \\
            &= \supp_X(U) \cup \supp_X(V) \\
            &\subseteq \{\vec x\} \cup \{\vec y\} \tag{IH}
        \end{align*}
    \end{itemize}
\end{proof}

The next result now tells us that \(\SL, \GA, \CA\) are all supported malleable.

\propMalleableSufficient*

Note that in a skew commutative skew associative theory, the reverse direction for skew associativity holds as well, since there must exist \(\sigma_1',\sigma_2', \sigma_1'', \sigma_2''\) below:
\begin{align*}
    \sigma_1(\sigma_2(x, y), z)
    &= \sigma_1'(z, \sigma_2(x, y)) \tag{skew commutative}\\
    &= \sigma_1'(z, \sigma_2'(y, x)) \tag{skew commutative}\\
    &= \sigma_1'(\sigma_2'(z, y),x) \tag{skew associative}\\
    &= \sigma_1''(x, \sigma_2'(z, y)) \tag{skew commutative}\\
    &= \sigma_1''(x, \sigma_2''(y, z)) \tag{skew commutative}
\end{align*}

\begin{proof}[Proof of \cref{prop:sufficient malleable}.]
    We are going to prove something a bit stronger.
    We are going to show that we can specifically take \(s(u,v) = c\) (a constant) or \(s(u,v) = \sigma(u, v)\) for some binary \(\sigma \in S\).
    To this end, we establish the following claim.
    We need a definition to properly state the claim:
    Call a sequence \((x_1, \dots, x_n)\) of variables and constants (allowing for repetitions) the \emph{appearance sequence} for a term \(t\) if the variables and constants appearing in \(t\) are precisely \(\{x_1,\dots, x_n\}\) and for any \(i < j\), \(x_i\) appears to the left of \(x_j\).
    For example, for operations \(\sigma_1,\sigma_2,\sigma_3 \in S\) and a constant \(c\), the term \(\sigma_1(x, \sigma_2(\sigma_3(c, y), z))\) has the appearance sequence \((x, c, y, z)\).
    It is worth noting that applying the skew associativity rule above does not change the appearance sequence of a term.
    Moreover, applying either rule maintains the length of the appearance sequence.
    \smallskip

    \noindent\textbf{Claim 1.} Let \(t \in S^* X\) with appearance sequence \((x_1, \dots, x_n)\).
    Then for any \(i < n\), there is a term \(t' \in S^*X\) such that \(\T \vdash t = t'\) and the appearance sequence for \(t'\) is \((x_1, \dots, x_{i+1}, x_i, \dots, x_n)\).\smallskip

    In other words, we can always ``nudge'' a variable to the left or right.\smallskip

    \noindent\emph{Proof of claim 1.}
    By induction on the length of the appearance sequence of \(t\).
    We cover two base cases: \(t = x_1\) and \(t = \sigma(x_1, x_2)\).
    In the first, the claim is vacuous.
    In the second, apply skew commutativity to obtain \(\T \vdash \sigma(x_1, x_{2}) = \sigma'(x_{2}, x_1)\) for some \(\sigma' \in S\).

    Now assume that the claim is true for any \(t'\) with an appearance sequence shorter than \((x_1, \dots, x_n)\).
    We need to show that the claim is true for a term of the form \(t = \sigma(t_1, t_2)\) whose appearance sequence is at least of length \(3\).
    There are three cases to consider.
    \begin{itemize}
        \item Suppose \(x_i\) and \(x_{i+1}\) both appear in \(t_1\) and let the appearance sequence of \(t_1\) be \((x_1, \dots, x_i, x_{i+1}, \dots, x_m)\).
        Then by the induction hypothesis, there is a term \(t_1'\) such that \(\T\vdash t_1 = t_1'\) and the appearance sequence of \(t_1'\) is \((x_1, \dots, x_{i+1}, x_i, \dots, x_m)\).
        We have \(\T \vdash t = \sigma(t_1, t_2) = \sigma(t_1', t_2)\) and the appearance sequence of \(\sigma(t_1', t_2)\) is \((x_1, \dots, x_{i+1}, x_i, \dots, x_m, x_{m+1}, \dots, x_n)\).

        \item Suppose \(x_i\) and \(x_{i+1}\) both appear in \(t_2\).
        Similar to the previous case.

        \item Suppose that the appearance sequences of \(t_1\) and \(t_2\) are \((x_1, \dots, x_i)\) and \((x_{i+1}, \dots, x_n)\) respectively.
        Then there are two subcases to consider:
        \begin{itemize}
            \item Suppose \(t_2 = \tau(t_2', t_2'')\) for some \(\tau \in S\), and the appearance sequences of \(t_2', t_2''\) are \((x_{i+1}, \dots, x_m)\) and \((x_{m+1}, \dots, x_n)\) respectively.
            Then there are \(\sigma',\tau',\tau'' \in S\) such that
            \begin{align*}
                \T \vdash \sigma(t_1, t_2)
                &= \sigma(t_1, \tau(t_2', t_2''))  \\
                &= \sigma'(\tau'(t_1, t_2'), t_2'') \tag{skew associativity}
            \end{align*}
            Now we can apply the induction hypothesis to \(\tau'(t_1, t_2')\) to obtain a term \(\T \vdash t'' = \tau'(t_1, t_2')\) such that the appearance sequence of \(t''\) is \((x_1, \dots, x_{i+1}, x_i, \dots, x_m)\).

            \item Suppose \(t_1 = \tau(t_1', t_1'')\) for some \(\tau \in S\), and the appearance sequences of \(t_1', t_1''\) are \((x_1, \dots, x_m)\) and \((x_{m+1}, \dots, x_i)\) respectively.
            Then there are \(\sigma',\tau',\tau'' \in S\) such that
            \begin{align*}
                \T \vdash \sigma(t_1, t_2)
                &= \sigma(\tau(t_1', t_1''), t_2)  \\
                &= \sigma'(t_1', \tau'(t_1'', t_2)) \tag{skew assoc.~in reverse}
            \end{align*}
            Now we can apply the induction hypothesis to \(\tau'(t_1'', t_2)\) to obtain a term \(\T \vdash t'' = \tau'(t_1'', t_2)\) such that the appearance sequence of \(t''\) is \((x_{m+1}, \dots, x_{i+1}, x_i, \dots, x_n)\).
        \end{itemize}
        This is an exhaustive case analysis because there are only binary operations and constants in \(S\) and the length of the appearance sequence of \(t\) is at least \(3\).
    \end{itemize}
    This concludes the proof of claim 1.

    \smallskip
    Now, consider a term \(\sigma(t_1, t_2)\) such that \((x_1, \dots, x_i)\) is the appearance sequence of \(t_1\).
    Then \(i\) is called the \emph{index of \(\sigma\) in \(t\)}.
    \smallskip

    \noindent\textbf{Claim 2.} Let \(t = \sigma(t_1, t_2)\) such that the appearance sequence of \(t\) is \((x_1, \dots, x_n)\) and the index of \(\sigma\) in \(t\) is \(i\).
    Then
    \begin{enumerate}
        \item if \(i + 1 < n\), there is a term \(t' = \sigma'(t_1', t_2')\) such that the appearance sequence of \(t'\) is \((x_1, \dots, x_n)\), \(\T \vdash t = t'\), and the index of \(\sigma'\) in \(t'\) is \(i + 1\); and
        \item if \(1 < i\), there is a term \(t' = \sigma'(t_1', t_2')\) such that the appearance sequence of \(t'\) is \((x_1, \dots, x_n)\), \(\T \vdash t = t'\), and the index of \(\sigma'\) in \(t'\) is \(i - 1\).
    \end{enumerate}

    \noindent\emph{Proof of claim 2.}
    We will argue for these two items simultaneously by induction on \(t\).
    There are two base cases, because of the restrictions on \(i\) in either situation (in particular, \(n > 2\)).
    The first base case for item 1 is \(t = \sigma(x_1, \tau(x_2, x_3))\).
    This is handled swiftly by skew associativity, because there are \(\sigma',\tau' \in S\) such that \(\T \vdash t = \sigma(x_1, \tau(x_2, x_3)) = \sigma'(\tau'(x_1, x_2), x_3)\).
    The index of \(\sigma\) in \(t\) goes from \(1\) to the index of \(\sigma'\), which is \(2\) here.
    The second base case is for item 2 and covers \(t = \sigma(\tau(x_1, x_2), x_3)\).
    This is similarly handled by skew associativity: there are \(\sigma',\tau' \in S\) such that \(\T \vdash t = \sigma(\tau(x_1, x_2), x_3) = \sigma'(x_1, \tau'(x_2, x_3))\).
    The index of \(\sigma\) goes from \(2\) to the index of \(\sigma'\), which is \(1\).

    For the induction step, we need only consider the following two cases.
    \begin{itemize}
        \item If the index of \(\sigma\) in \(t\) is \(i < n-1\), then there exist \(t_1,t_2,t_3\) and \(\tau\in S\) such that \(t = \sigma(t_1, \tau(t_2, t_3))\).
        By repeatedly applying the induction hypothesis, item 2, to \(\tau(t_2, t_3)\), we can find a \(\tau'\) and \(t_3'\) such that \(\T \vdash \tau(t_2, t_3) = \tau'(x_{i+1}, t_3')\) and the conditions of item 2 are satisfied.
        Then we can manipulate terms
        \begin{align*}
            \T \vdash t
            &= \sigma(t_1, \tau(t_2, t_3)) \\
            &= \sigma(t_1, \tau'(x_{i+1}, t_3')) \\
            &= \sigma'(\tau''(t_1, x_{i+1}), t_3') \tag{skew associativity}
        \end{align*}
        Thus, the index of \(\sigma'\) in \(t' = \sigma'(\tau''(t_1, x_{i+1}), t_3')\) is \(i + 1\).

        \item If the index of \(\sigma\) in \(t\) is \(1 < i\), then there exist \(t_1,t_2,t_3\) and \(\tau\in S\) such that \(t = \sigma(\tau(t_1, t_2), t_3)\).
        By repeatedly applying the induction hypothesis, item 1, to \(\tau(t_1, t_2)\), we can find a \(\tau'\) and \(t_1'\) such that \(\T \vdash \tau(t_1, t_2) = \tau'(t_1', x_i)\) and the conditions of item 1 are satisfied.
        Then we can manipulate terms
        \begin{align*}
            \T \vdash t
            &= \sigma(\tau(t_1, t_2), t_3) \\
            &= \sigma(\tau'(t_1', x_i), t_3) \\
            &= \sigma'(t_1', \tau''(x_i, t_3)) \tag{skew associativity}
        \end{align*}
        Thus, the index of \(\sigma'\) in \(t' = \sigma'(t_1', \tau''(x_i, t_3))\) is \(i - 1\).
    \end{itemize}
    This concludes the proof of Claim 2.

    Now let us turn to the result we are ultimately aiming at.
    If \(t = x\) or \(t = c\), there is nothing to prove.
    Otherwise, given a term \(t \in S^*X\) with appearance sequence \((x_1, \dots, x_n)\), \(n > 1\), and a partition \(X = U + V\), we proceed as follows.
    If \(|U \cap \{x_1, \dots, x_n\}| = m\), repeatedly apply Claim 1 to obtain a term \(t'\) with appearance sequence \((u_1, \dots, u_m, v_{m+1}, \dots, v_n)\) such that \(u_i \in U\), \(v_j \in V\), and \(\{x_1, \dots, x_n\} = \{u_1, \dots, u_m, v_{m + 1}, \dots, v_n\}\), and such that \(\T \vdash t = t'\).
    Let \(t' = \sigma(t_1, t_2)\).
    If \(t_1 = t_1(u_1, \dots, u_m), t_2 = t_2(v_{m+1}, \dots, v_{n})\), then take \(s(u, v) = \sigma(u, v)\).
    Otherwise, repeatedly apply Claim 2 to obtain \(t'' = \sigma'(t_1', t_2')\) such that \(\T \vdash t' = t''\), the appearance sequence of \(t''\) is \((u_1, \dots, u_m, v_{m+1}, \dots, v_n)\), and the index of \(\sigma'\) in \(t''\) is \(m\).
    Then \(t'' = \sigma'(t_1(\vec u), t_2'(\vec v))\), so take \(s(u,v) = \sigma'(u, v)\).
\end{proof}

The theories \(\SL, \GA, \CA\) are all skew commutative and skew associative, so \cref{prop:sufficient malleable} tells us thay are supported malleable.
In fact, as noted in the proof, every binary star in any of the three theories is equivalent to one of the form \(s^{(\sigma)}\), where \(\sigma\) is a binary operation (and not a more complex term).

As we saw in the main text, \(\GC\) is not skew-associative, so we need to prove that \(\GC\) is malleable separately.

\begin{proposition}
    The equational theory \(\GC\) is malleable.
\end{proposition}

Given \(\theta \in \D X\) and \(U \subseteq X\), define \(\theta(U) = \sum_{x \in U} \theta(x)\).

\begin{proof}
    We sketch the proof as follows.
    Let \(\chi \colon \At \to \D(\bot + X)\) and \(X = U + V\).
    For each \(\alpha \in \At\), define the following quantities.
    \begin{gather*}
        r_\alpha = \chi(\alpha)(X)
        \qquad
        p_\alpha = \frac{1}{r_\alpha} \chi(\alpha)(U)
    \end{gather*}
    Above, \(p_\alpha\) is only defined if \(r_\alpha > 0\).
    This allows us to define \(s(u,v)\) as follows.
    Let \(\At = \{\alpha_1, \dots, \alpha_n\}\).
    Then
    \[
        s(u, v)^\rho(\alpha) = \begin{cases}
            (u \oplus_{p_\alpha}^\rho v) \oplus_{r_\alpha}^\rho 0 & r_\alpha > 0, 0 < p_\alpha < 1  \\
            u \oplus_{r_\alpha}^\rho 0 & r_\alpha > 0, p_\alpha = 1 \\
            v \oplus_{r_\alpha}^\rho 0 & r_\alpha > 0, p_\alpha = 0 \\
            0 & r_\alpha = 0
        \end{cases}
    \]
    For each \(\alpha \in \At\) such that \(r_\alpha > 0\) and \(0 < p_\alpha < 1\), define
    \[
        t_1^\rho(\alpha)(\xi) = \begin{cases}
            \frac{1}{r_\alpha p_\alpha} \chi(\alpha)(\xi) & \xi \in U \\
            0 & \xi \in V \\
            1 - r_\alpha & \xi = \bot
        \end{cases}
        \qquad
        t_2^\rho(\alpha)(\xi) = \begin{cases}
            \frac{1}{r_\alpha (1 - p_\alpha)} \chi(\alpha)(\xi) & \xi \in V \\
            0 & \xi \in U \\
            1 - r_\alpha & \xi = \bot
        \end{cases}
    \]
    And similarly, if \(r_\alpha > 0\) and \(p_\alpha = 1\), define
    \[
        t_1^\rho(\alpha)(\xi) = \begin{cases}
            \frac{1}{r_\alpha} \chi(\alpha)(\xi) & \xi \in U \\
            0 & \xi \in V \\
            1 - r_\alpha & \xi = \bot
        \end{cases}
        \qquad
        t_2^\rho(\alpha)(\xi) = \begin{cases}
            0 & \xi \in V \\
            0 & \xi \in U \\
            1 & \xi = \bot
        \end{cases}
    \]
    And if \(r_\alpha > 0\) and \(p_\alpha = 0\), define
    \[
        t_1^\rho(\alpha)(\xi) = \begin{cases}
            0 & \xi \in U \\
            0 & \xi \in V \\
            1 & \xi = \bot
        \end{cases}
        \qquad
        t_2^\rho(\alpha)(\xi) = \begin{cases}
            \frac{1}{r_\alpha} \chi(\alpha)(\xi) & \xi \in V \\
            0 & \xi \in U \\
            1 - r_\alpha & \xi = \bot
        \end{cases}
    \]
    Then by design, \(t_1^\rho \in \D(\bot + U)\), \(t_2^\rho \in \D(\bot + V)\), and \(\chi = s^\rho(t_1^\rho, t_2^\rho)\).
\end{proof}